\documentclass[fleqn,usenatbib]{mnras}

\usepackage{newtxtext,newtxmath}

\usepackage[T1]{fontenc}

\DeclareRobustCommand{\VAN}[3]{#2}
\let\VANthebibliography\thebibliography
\def\thebibliography{\DeclareRobustCommand{\VAN}[3]{##3}\VANthebibliography}

\usepackage{graphicx}	
\usepackage{amsmath}	
\usepackage{CJKutf8} 

\usepackage{bm}
\usepackage{array}

\newcommand{\be}{\begin{equation}}
\newcommand{\ee}{\end{equation}}
\newcommand{\bea}{\begin{eqnarray}}
\newcommand{\eea}{\end{eqnarray}}
\newcommand{\dd}{\mathrm{d}}

\title[The molecular disc scale-height]{On the scale-height of the molecular gas disc in Milky Way-like galaxies}

\author[Jeffreson, Sun \& Wilson]{
	Sarah M.~R.~Jeffreson$^{1}$, Jiayi Sun (\begin{CJK}{UTF8}{gbsn} 孙嘉懿 \end{CJK})$^{2,3,4,*}$ and Christine D.~Wilson$^{2}$
	\\
$^{1}$ Center for Astrophysics, Harvard \& Smithsonian, 60 Garden St, Cambridge, MA 02138, USA \\
$^{2}$ Department of Physics and Astronomy, McMaster University, 1280 Main St. West, Hamilton ON L8S 4M1 Canada \\
$^{3}$ Canadian Institute for Theoretical Astrophysics (CITA), University of Toronto, 60 St George Street, Toronto, ON M5S 3H8, Canada \\
$^{4}$ Department of Astronomy, The Ohio State University, 140 West 18th Avenue, Columbus, OH 43210, USA \\
$^*$ CITA National Fellow \\
}

\date{Accepted XXX. Received XXX; in original form XXX}

\pubyear{2021}

\begin{document}
\label{firstpage}
\pagerange{\pageref{firstpage}--\pageref{lastpage}}
\maketitle

\begin{abstract}
We study the relationship between the scale-height of the molecular gas disc and the turbulent velocity dispersion of the molecular interstellar medium within a simulation of a Milky Way-like galaxy in the moving-mesh code {\sc Arepo}. We find that the vertical distribution of molecular gas can be described by a Gaussian function with a uniform scale-height of $\sim 50$~pc. We investigate whether this scale-height is consistent with a state of hydrostatic balance between gravity and turbulent pressure. We find that the hydrostatic prediction using the total turbulent velocity dispersion (as one would measure from kpc-scale observations) gives an over-estimate of the true molecular disc scale-height. The hydrostatic prediction using the velocity dispersion between the centroids of discrete giant molecular clouds (cloud-cloud velocity dispersion) leads to more-accurate estimates. The velocity dispersion internal to molecular clouds is elevated by the locally-enhanced gravitational field. Our results suggest that observations of molecular gas need to reach the scale of individual molecular clouds in order to accurately determine the molecular disc scale-height.
\end{abstract}



\section{Introduction} \label{Sec::Introduction}
The vertical distribution of gas in external disc galaxies is the key piece of information required to connect the observed properties of galactic-scale star formation to the physics driving these trends. That is, the slope and normalisation of the three-dimensional relationship between the star formation rate volume density $\rho_{\rm SFR}$ and the gas volume density $\rho$, averaged across kiloparsec-scale (kpc-scale) regions of the galactic disc, varies between competing theories of galactic-scale star formation~\citep[e.g.][]{Tan00,2000ApJ...530..277E,Krumholz+McKee05,Ostriker+10,Semenov18,Jeffreson+Kruijssen18,2022arXiv220600681O}. To falsify these theories, i.e.~to distinguish which theories best reproduce the slope and normalisation of the observed two-dimensional relationship between the kpc-scale star formation rate surface density $\Sigma_{\rm SFR}$ and gas surface density $\Sigma$~\citep[e.g.][]{Kennicutt98,WongBlitz2002,Bigiel08,Leroy+09,Saintonge+11,Schruba11,Saintonge+11b,Leroy+13}, a gas distribution along the line-of-sight must be known (or assumed) for external galaxies at face-on or inclined viewing angles. In many galaxies, the scale-height of the gas disc is also similar to the transition point between the two- and three-dimensional forms of the turbulence power spectrum~\citep{2004RvMP...76..125M}. It may therefore encode information about the physics that drive turbulence in the interstellar medium.

The vertical distribution of diffuse, atomic gas in massive disc galaxies is commonly assumed to result from a state of hydrostatic equilibrium, in which the gravitational force due to the combined potential of dark matter, stars and gas balances the effective pressure of the gas, due to its combined turbulent and thermal velocity dispersion. Hydrostatic equilibrium is assumed in calculations of a wide range of galaxy properties, including the shape of the Milky Way's dark matter halo~\citep{2000MNRAS.311..361O}, the mid-plane pressure in nearby galaxies~\citep{Elmegreen89,Elmegreen93} and its connection to the ratio of molecular to atomic gas~\citep{BlitzRosolowsky2004,2006ApJ...650..933B,2008AstL...34..152K,2011AJ....141...48Y,2014AJ....148..127Y}, as well as the relationship between giant molecular cloud properties and the ambient pressure of gas at the galactic mid-plane~\citep[e.g.][]{2011MNRAS.416..710F,Hughes13b,2019ApJ...883....2S,Sun2020,2020MNRAS.498..385J}. It also underpins feedback-driven models of self-regulated star formation~\citep[e.g.][]{2005ApJ...630..167T,Ostriker+10,OstrikerShetty2011,2013MNRAS.433.1970F,Krumholz18b}.

Direct observational evidence for hydrostatic equilibrium in atomic gas discs is limited, as observations of external galaxies yield \textit{either} the vertical profile of the atomic gas distribution \citep[edge-on viewing angle, e.g.][]{2011AJ....141...48Y,2014AJ....148..127Y} \textit{or} the radial profile of the vertical gas velocity dispersion \citep[face-on viewing angle, e.g.][]{Tamburro09}. Confirmation of hydrostatic equilibrium requires the comparison of these two datasets. However, three-dimensional numerical simulations have provided substantial evidence that the hydrostatic condition yields reasonable estimates for the atomic disc scale-height and galactic mid-plane pressure on kpc scales~\citep[e.g.][]{2009ApJ...693.1346K,2011ApJ...743...25K,2013ApJ...776....1K,KimCG&Ostriker15b,2022arXiv220600681O}, so long as the disc is in the steady-state (e.g.~the gas discs in massive, isolated galaxies without prominent outflows). The atomic disc scale-height in the Milky Way~\citep[][]{1995ApJ...448..138M} is also roughly consistent with the hydrostatic prediction if a constant atomic gas velocity dispersion of 8 km/s~\citep{Spitzer78} is assumed at all galactocentric radii~\citep{2002A&A...394...89N}.

\begin{figure*}
\includegraphics[width=\linewidth]{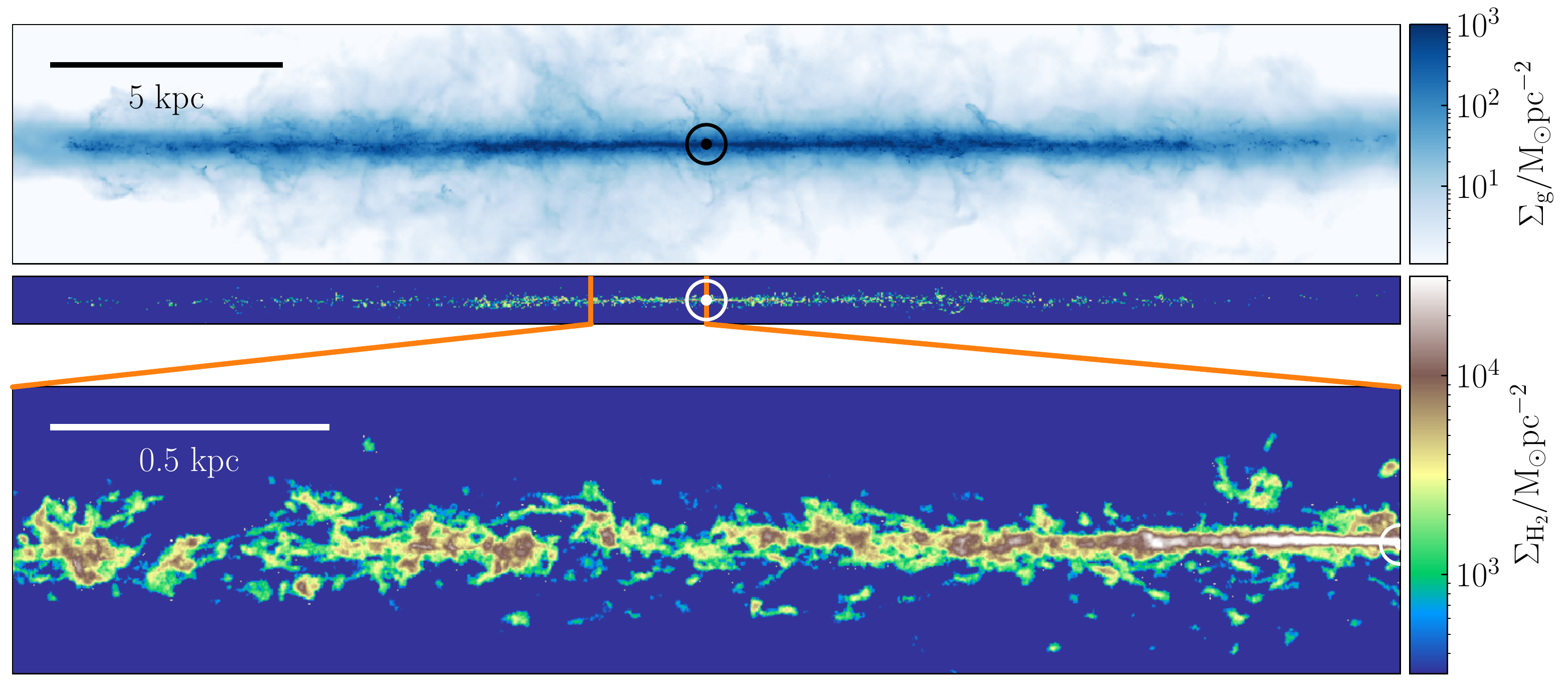}
\caption{Column density maps of the total (upper panel) and molecular (lower two panels) gas distribution for the simulated galaxy, viewed across the galactic mid-plane. The galactic centre is marked by a circled point in each panel.}
\label{Fig::morphology}
\end{figure*}

Recently, the assumption of hydrostatic equilibrium has also been applied to calculate the galactic-scale vertical distribution of the molecular gas in galaxies~\citep{2019A&A...622A..64B,2019ApJ...882....5W,2021MNRAS.501.3527P}. However, the molecular gas is denser, clumpier, and has locally-enhanced self-gravity in the dense regions~\citep{Kennicutt+Evans12}. In this case, a significant portion of the gas motions within the dense gas clouds arises in response to the locally-enhanced self-gravity, putting these clouds in an `overpressured' state relative to the ambient ISM~\citep[e.g.][]{Sun2020,2020MNRAS.498..385J}. This introduces some challenges to determining the overall scale height of the molecular gas disc from a naive hydrostatic calculation. In particular, the scale height then includes two components, one from the vertical extent of the largest clouds, and the other from the vertical distribution of the cloud centroids relative to the mid-plane. Measurements of cloud sizes and vertical distribution in the Milky Way suggest that the latter component is dominant~\citep{2015ARA&A..53..583H}. However, no empirical estimates of this component exist for external galaxies to our knowledge, and the methodological framework for deriving such estimates is yet to be constructed.

In this work, we lay out a framework for estimating the vertical scale-height of the molecular gas disc from a refined hydrostatic calculation, taking into account the clumpiness of the molecular gas and distinguishing the different components of its velocity dispersion on different spatial scales. We then test the fidelity of this framework using a realistic simulation of a Milky Way-like galaxy in the moving-mesh code {\sc Arepo}, and discuss its applicability to state-of-the-art observations of external galaxies. In Section~\ref{Sec::sims} we re-iterate the key details of the isolated galaxy simulation, first presented in~\cite{2021MNRAS.505.3470J}. In Section~\ref{Sec::theory} we derive the molecular disc scale-height from the condition for hydrostatic equilibrium. Section~\ref{Sec::sh-sims} details the calculation of this scale-height from the observable properties of our simulated galaxy, and compares it to the true three-dimensional scale-height. Section~\ref{Sec::discussion} presents a discussion of our results in the context of previous simulations and observations of hydrostatic equilibrium in disc galaxies. Finally, a summary of our conclusions is given in Section~\ref{Sec::conclusions}.

\section{Simulations} \label{Sec::sims}
We analyse the simulation of a Milky Way-like galaxy first presented in~\cite{2021MNRAS.505.3470J} (named `HII heat and beamed mom' in that work). The spatial distribution of the total (upper panel) and molecular (lower panels) gas reservoirs is shown at an edge-on viewing angle in Figure~\ref{Fig::morphology}. As discussed in~\cite{2021MNRAS.505.3470J}, the simulated galaxy reproduces the key observable large-scale, intermediate-scale and cloud-scale properties of Milky Way-like galaxies, including the radial profiles of the gas column density and velocity dispersion~\citep{Tamburro09,Yin09}, the relation of the star formation rate column density to the total and molecular gas column densities~\citep[e.g.][]{Kennicutt98,Bigiel08}, the phase structure of the interstellar medium~\citep[e.g.][]{Wolfire03}, and the mass and size distributions of its giant molecular clouds~\citep[e.g.][]{Rosolowsky03,Freeman17,Miville-Deschenes17,Colombo+19}. The snapshot analysed in this work is taken at a simulation time of $600$~Myr. It has a gas-to-stellar mass ratio of 0.13, with around 40~per~cent of gas in the molecular phase, by mass. Here we give an overview of the most important features of our numerical method, and refer the reader to~\cite{2021MNRAS.505.3470J} for a more complete explanation.

The simulation is run using the {\sc Arepo} code~\citep{Springel10}. The gravitational acceleration vectors for all gas, stellar and dark matter particles are computed using a hybrid TreePM gravity solver. The gaseous (hydrodynamical) component is modelled by the unstructured moving mesh defined by the Voronoi tessellation about a discrete set of points that move according to the local gas velocity. We evolve the isolated Milky Way-like initial condition generated for the Agora comparison project~\citep{Kim14} at a mass resolution of $1.254 \times 10^7~{\rm M}_\odot$ for the dark matter particles, $3.437 \times 10^5~{\rm M}_\odot$ for the stellar particles, and $859~{\rm M}_\odot$ for the gas cells. The dark matter halo follows the profile of~\cite{Navarro97} with concentration parameter $c=10$, spin parameter $\lambda=0.04$, mass $M_{200}=1.07 \times 10^{12}~{\rm M}_\odot$, and virial radius $R_{200} = 205~{\rm kpc}$. The initial condition has a~\cite{Hernquist90} stellar bulge of mass $3.437 \times 10^9~{\rm M}_\odot$ and an exponential disc of mass $4.297 \times 10^{10}~{\rm M}_\odot$, with scale-length $3.43~{\rm kpc}$ and scale-height $0.34~{\rm kpc}$. The bulge to stellar disc ratio is $0.125$ and the initial ratio of gas to stellar mass is $0.18$. During run-time, we employ the adaptive gravitational softening scheme in {\sc Arepo} with a softening length of $1.5$ times the Voronoi gas cell size, with a minimum value of $20$~pc. We set a softening length of $20$~pc for the stellar particles, and a softening length of $260$~pc for the dark matter particles, according to the convergence tests presented in~\cite{2003MNRAS.338...14P}. Given that the gas disc scale-height and Toomre mass are resolved at all scales in our simulations, the adaptive softening scheme ensures that minimal artificial fragmentation occurs at scales larger than the Jeans length~\citep{Nelson06}.

The initial gas temperature in our simulation is set to $10^4~{\rm K}$, and this re-equilibriates on a time-scale of $\la 10~{\rm Myr}$ to a state of thermal balance between line-emission cooling and heating due to photo-electric emission from dust grains and polycyclic aromatic hydrocarbons, modelled using the simplified network of hydrogen, carbon and oxygen chemistry introduced in~\cite{NelsonLanger97,GloverMacLow07a,GloverMacLow07b}. For each gas cell, the network computes and tracks fractional abundances for the species ${\rm H}$, ${\rm H}_2$, ${\rm H}^+$, ${\rm He}$, ${\rm C}^+$, ${\rm CO}$, ${\rm O}$ and ${\rm e}^-$. This chemistry is self-consistently coupled to the heating and cooling of the interstellar medium, according to the atomic and molecular cooling function of~\cite{Glover10}. These authors give the full list of included heating and cooling processes in their Table 1. The thermal evolution of the gas in our simulations therefore depends on the gas density and temperature, as well as on the strength of the interstellar radiation field, the cosmic-ray ionisation rate, the dust fraction and temperature, and on the set of chemical abundances tracked for each gas cell. We take a value of $1.7$~Habing fields for the UV component of the ISRF~\citep{Mathis83} and a value of $3 \times 10^{-17}$~s$^{-1}$ to the cosmic ionisation rate~\citep{2000A&A...358L..79V}. We assume the solar neighbourhood value of the dust-to-gas ratio.

Our star formation prescription locally reproduces the observed relation of~\cite{Kennicutt98} between the SFR surface density and the gas surface density, following the equation
\begin{equation}
\label{Eqn::starformation}
\frac{\dd \rho_{*,i}}{\dd t} = 
\begin{cases}
      \frac{\epsilon_{\rm ff} \rho_i}{t_{{\rm ff},i}}, \; \rho_i \geq \rho_{\rm thres} \\
      0, \; \rho_i < \rho_{\rm thres}\\
  \end{cases},
\end{equation}
where $t_{{\rm ff}, i} = \sqrt{3\pi/(32 G\rho_i)}$ is the local free-fall time-scale for a mass volume density of $\rho_i$. The star formation efficiency per free-fall time is set to $\epsilon_{\rm ff} = 1$~per~cent, according to the measured gas depletion time across nearby galaxies~\citep{Leroy17,Krumholz&Tan07,Krumholz18,Utomo18}. A local star formation density threshold of $\rho_{\rm thresh} = 1000~{\rm cm}^{-3}$ is used to ensure that the densest gas in each simulation is Jeans-unstable at the median mass resolution of $859~{\rm M}_\odot$ (assuming that the star-forming gas has a maximum temperature of $100$~K and is in approximate thermal equilibrium). Each resulting star particle is assigned a stellar population drawn stochastically from a~\cite{Chabrier03} initial stellar mass function (IMF) via the Stochastically Lighting Up Galaxies (SLUG) stellar population synthesis model~\citep{daSilva12,daSilva14,Krumholz15}. By evolving each stellar population along Padova solar metallicity tracks~\cite{Fagotto94a,Fagotto94b,VazquezLeitherer05}, using {\sc Starburst99}-like spectral synthesis~\citep{Leitherer99}, SLUG provides an ionising luminosity for each star particle at each simulation time-step, as well as the number of supernovae $N_{*, {\rm SN}}$ it has generated and the mass $\Delta m_*$ it has ejected.

Using the values of $N_{*, {\rm SN}}$ and $\Delta m_*$ provided for each star particle by our stellar evolution model, we model the momentum and thermal energy injected by supernova explosions at each simulation time-step. If $N_{*, {\rm SN}} = 0$, then we assume that any mass loss results from stellar winds. If $N_{*, {\rm SN}} > 0$, we assume that all mass loss results from supernovae. Our simulations do not resolve the energy-conserving/momentum-generating phase of supernova blast-wave expansion, so we explicitly inject the terminal momentum of the blast-wave to avoid over-cooling, as discussed in~\cite{KimmCen14}. We use the unclustered parametrisation of the terminal momentum injected into the gas cells $k$ that share faces with a central cell $j$, derived from the high-resolution simulations of~\cite{Gentry17}, and given by
\begin{equation} \label{Eqn::Gentry17}
\frac{p_{{\rm t}, k}}{{\rm M}_\odot {\rm kms}^{-1}} = 4.249 \times 10^5 N_{j, {\rm SN}} \Big(\frac{n_k}{{\rm cm}^{-3}}\Big)^{-0.06},
\end{equation}
where $N_{j, {\rm SN}}$ is the total number of supernovae associated with all star particles for which gas cell $j$ is the nearest neighbour. We distribute this terminal momentum to the gas cells surrounding the central cell, as described in~\cite{2020arXiv200403608K,2021MNRAS.505.3470J}. The upper limit on the terminal momentum is set by kinetic energy conservation as the shell sweeps through the gas cells $k$~\citep[see also][for similar prescriptions]{Hopkins18,Smith2018}.

In addition to supernova feedback, we include pre-supernova feedback from HII regions, following~\cite{2021MNRAS.505.3470J}. This model takes account of the momentum injected by both radiation pressure and by the thermal pressure from heated gas inside the HII region, according to the analytic work of~\cite{Matzner02,KrumholzMatzner09}. We group the star particles in the simulation via a Friends-of-Friends prescription of linking length equal to the HII region ionisation front radius, improving the numerical convergence of the feedback model. The momentum is distributed to the gas cells that adjoin a central cell closest to the centre of luminosity of each Friends-of-Friends group. The gas cells inside the resulting grouped Str\"{o}mgren radii are also heated self-consistently and held above a temperature floor of $7000$~K, for as long as they receive ionising photons from the star particles. We rely on the chemical network to ionise the gas in accordance with the thermal energy injected, and so do not explicitly adjust the chemical state of the heated gas cells.

\begin{figure}
\includegraphics[width=\linewidth]{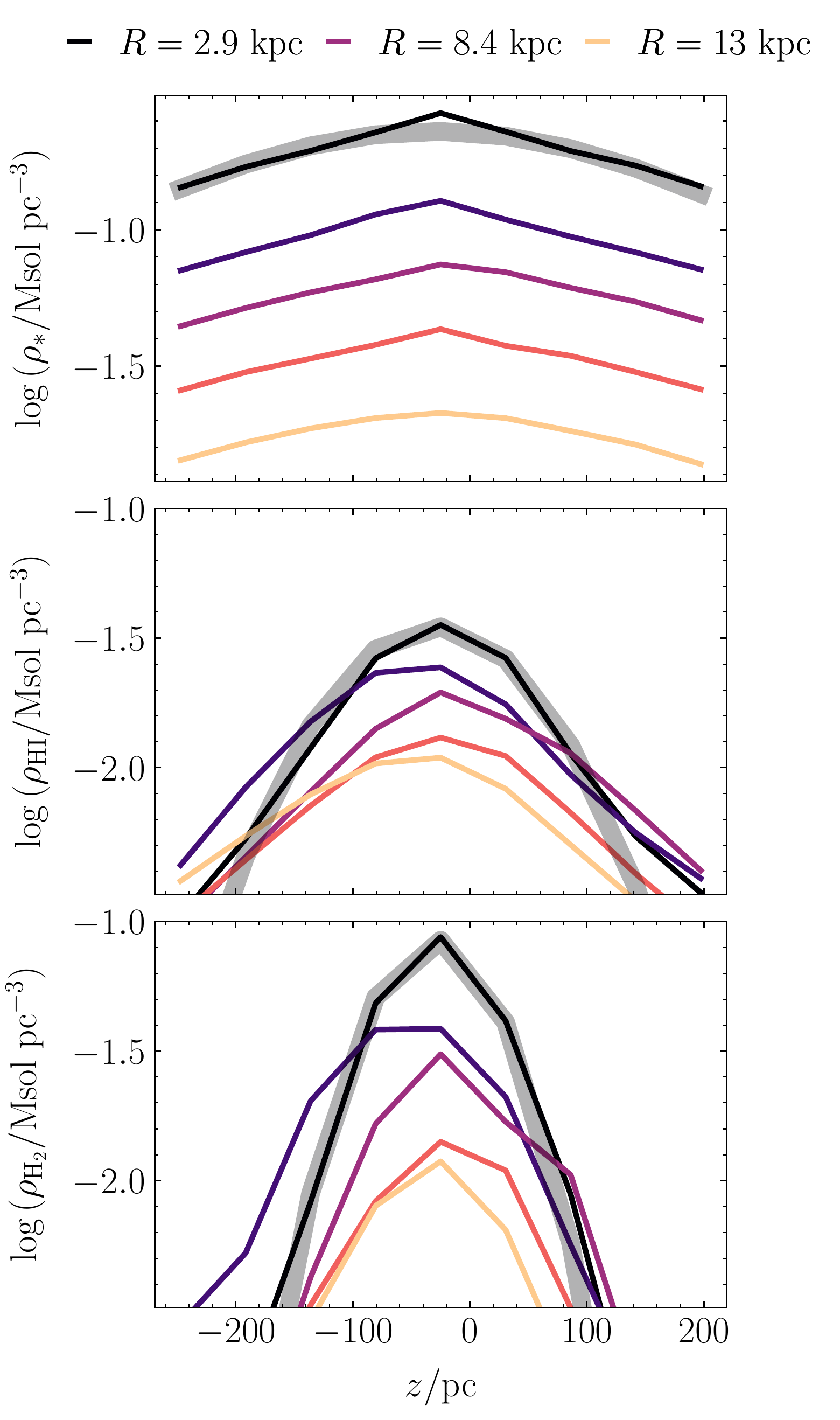}
\caption{The vertical profiles of the molecular gas volume density $\rho_{\rm H_2}$ (lower panel) and the atomic gas volume density $\rho_{\rm HI}$ (centre panel) in five different kpc-scale bins of galactocentric radius $R$. Lighter line colours correspond to larger galactocentric radii. These gas profiles follow approximately-Gaussian distributions, in accordance with the hydrostatic condition given by Equation~(\ref{Eqn::rho-v2}). The stellar volume density (upper panel) is well-described by a hyperbolic secant profile. The thick grey lines show analytic Gaussian and hyperbolic secant profiles, to guide the eye.}
\label{Fig::z-rho-dstbn}
\end{figure}

\section{Theory} \label{Sec::theory}
In this section, we derive an analytic prediction for the scale-height of an axisymmetric gas disc in a state of hydrostatic equilibrium.\footnote{We use the word `hydrostatic' purely in reference to the standard `hydrostatic equation', quantifying the state of balance between a gravitational potential gradient and a gradient in the (effective) pressure. We emphasise that the interstellar medium is dynamic and rapidly-evolving by nature. In this case, the `equilibrium' described by the hydrostatic equation applies only to an average over large spatial scales and long time-scales.} In this scenario, the gravitational force per unit volume that pulls the gas towards the galactic mid-plane is balanced by a vertical (effective) pressure gradient on galactic scales. The effective pressure includes only those random (thermal plus turbulent) motions that prevent the particles in the disc from simply falling towards the mid-plane, so provides an effective support term to counteract gravity. In Section~\ref{Sec::smooth-hyd} we consider the case of a gas-disc that is smooth on kpc-scales. In Section~\ref{Sec::clumpy-hyd} we consider the case of a clumpy gas disc, consisting of a population of discrete gas clouds.

\subsection{A smooth axisymmetric disc in hydrostatic equilibrium} \label{Sec::smooth-hyd}
For an axisymmetric disc in hydrostatic equilibrium,
\begin{equation} \label{Eqn::hyd-eqm-general}
\frac{\partial P}{\partial z} = -\rho \frac{\partial \Phi}{\partial z},
\end{equation}
where $P$ is the effective pressure, $\rho$ is the gas volume density and $\Phi$ is the gravitational potential. If the disc is cylindrically-rotating, and has a vertical turbulent velocity dispersion $\sigma_z$ that is supersonic and independent of the displacement $z$ from the galactic mid-plane, then the vertical effective pressure can be written as $P_z = \rho \sigma_z^2$. Substituting into Equation~(\ref{Eqn::hyd-eqm-general}) gives the integrated equation of motion
\begin{equation} \label{Eqn::hyd-eqm}
\frac{\partial \rho(R,z) \sigma_z^2(R)}{\partial z} = -\rho(R,z) \frac{\partial \Phi(R,z)}{\partial z},
\end{equation}
where $(R,\phi,z)$ are a set of cylindrical polar co-ordinates with their origin at the galactic centre. Integrating Equation~(\ref{Eqn::hyd-eqm}) over $z$ to obtain the functional form of the volume density gives
\begin{equation} \label{Eqn::rho-v1}
\rho(R,z) = \mathcal{C}_1(R) \exp{\Big[-\frac{\Phi(R,z)}{\sigma_z^2(R)}\Big]},
\end{equation}
where $\mathcal{C}_1(R)$ is a function independent of $z$.

We now follow a similar procedure to~\cite{1995AJ....110..591O,2002A&A...394...89N,2009ApJ...693.1346K,2019A&A...622A..64B} and express the gravitational potential as a Maclaurin series expansion about the galactic mid-plane, such that
\begin{equation} \label{Eqn::taylor-series}
\Phi(R,z) = \Phi(R,0) + \frac{z^2}{2} \frac{\partial^2 \Phi}{\partial z^2}\Big|_{z=0}(R) + \mathcal{O}\Big(\frac{z^4}{4!} \frac{\partial^4 \Phi}{\partial z^4}\Big|_{z=0}\Big).
\end{equation}
We have assumed that the matter distribution is symmetric about the mid-plane, so that the gravitational force there is zero, allowing us to set all odd derivatives of $\Phi$ to zero. If the total matter distribution varies slowly about the galactic mid-plane (not too heavily-peaked), then the final term in Equation~(\ref{Eqn::taylor-series}) becomes $\mathcal{O}(z^3/R^3)$ for a typical thin-disc potential~\citep[e.g.][]{Miyamoto&Nagai75}. In this case, we may truncate the series after the second derivative, so long as the gas under consideration is close to the galactic mid-plane ($z \ll R$). Substituting Equation~(\ref{Eqn::taylor-series}) back into Equation~(\ref{Eqn::rho-v1}) then shows that the distribution of gas volume densities about the galactic mid-plane is approximately Gaussian\footnote{In practice, the two approximations we have applied (a slowly-varying matter density distribution at the mid-plane, plus consideration of only the gas close to the mid-plane) mean that the total matter distribution must be vertically extended relative to the gas disc under consideration, in order for the predicted profile to be Gaussian. For the molecular distribution to be Gaussian, we must have $\rho_* + \rho_{\rm HI} > \rho_{\rm H_2}$.}, such that
\begin{equation} \label{Eqn::rho-v2}
\rho(R,z) = \rho(R,0) \exp{\Big[-\frac{z^2}{2h^2}\Big]},
\end{equation}
where $h$ is the disc scale-height, with a value defined by
\begin{equation} \label{Eqn::sh}
h^2 = \sigma_z^2(R) \Big[\frac{\partial^2 \Phi}{\partial z^2}\Big|_{z=0}(R)\Big]^{-1},
\end{equation}
equivalent to the standard deviation of the gas distribution. In Figure~\ref{Fig::z-rho-dstbn}, we show that the vertical distributions of molecular and atomic gas in our simulated disc galaxy both follow an approximately-Gaussian form at galactocentric radii $2~{\rm kpc} < R < 13~{\rm kpc}$, consistent with our assumptions so far.

We also note that the vertical stellar distribution in our simulated galaxy closely follows the hyperbolic secant form of~\cite{1942ApJ....95..329S}, as illustrated by the thick grey line in the upper panel of Figure~\ref{Fig::z-rho-dstbn}. This means that the mid-plane stellar volume density can be expressed in terms of the stellar surface density $\Sigma_*$ and scale-height $h_*$ according to $\rho_*|_{z=0} = \frac{\Sigma_*}{4h_*}$. In observational studies, the value of $h_*$ can in turn be approximated either by assuming a standard functional relation to the gas disc scale-length~\citep[e.g.][]{2008AJ....136.2782L,Sun2020}, or by solving a separate hydrostatic equation for the stellar disc (if integral field unit observations are present).

We may now write the disc scale-height in terms of the observable properties of the galaxy, using the Poisson equation
\begin{equation} \label{Eqn::Poisson-general}
\nabla^2 \Phi = 4\pi G \rho_{\rm total},
\end{equation}
where $G$ is the gravitational constant and $\rho_{\rm total}$ is the combined volume density of all galactic disc components (all gas phases and stars). For an axisymmetric disc, we can evaluate Equation~(\ref{Eqn::Poisson-general}) at the galactic mid-plane to give
\begin{equation} \label{Eqn::Poisson}
\frac{1}{R} \frac{\partial}{\partial R} \Big(R \frac{\partial \Phi}{\partial R}\Big|_{z=0}\Big) + \frac{\partial^2 \Phi}{\partial z^2} \Big|_{z=0} = 4\pi G \Big(\rho + \sum_i{\rho_i}\Big)\Big|_{z=0}.
\end{equation}
We have written $\rho_{\rm total} = \rho + \sum_i{\rho_i}$ in terms of the volume density $\rho$ of the gas phase under consideration, as well as the volume densities $\sum_i{\rho_i}$ of all disc components that do not contribute to $\rho$. For the molecular gas disc we have $\rho = \rho_{\rm H_2}$, so $\sum_i{\rho_i} = \rho_{\rm HI} + \rho_*$, where $\rho_{\rm HI}$ is the mid-plane atomic gas volume density and $\rho_*$ is the mid-plane stellar volume density. The first term on the left-hand side depends only on the galactic rotation curve $v_{\rm c}(R) = \sqrt{R\:\dd \Phi/\dd R}$, and can be written in terms of the orbital angular velocity $\Omega = v_{\rm c}/R$ and the galactic shear parameter $\beta = \dd \ln{v_{\rm c}}/\dd \ln{R}$. The shear parameter quantifies the degree of differential rotation between adjacent annuli within the disc (the degree of shearing increases from solid body rotation at $\beta = 1$ through to the case of a flat rotation curve at $\beta = 0$). Equation~(\ref{Eqn::Poisson}) becomes
\begin{equation}
\frac{\partial^2 \Phi}{\partial z^2}\Big|_{z=0} = \frac{2\sqrt{2\pi}G \Sigma(R)}{h} + 4\pi G\Big(\sum_i{\rho_i}\Big)\Big|_{z=0} -2\beta \Omega^2,
\end{equation}
where we have now integrated Equation~(\ref{Eqn::rho-v2}) over $z$ to write the mid-plane gas volume density as $\rho|_{z=0} = \Sigma/\sqrt{2\pi}h$. Finally, substituting this into Equation~(\ref{Eqn::sh}) gives
\begin{equation} \label{Eqn::sh-v2}
h^2 = \sigma^2_z(R) \Big[\frac{2\sqrt{2\pi}G\Sigma(R)}{h} + 4\pi G\Big(\sum_i{\rho_i}\Big)\Big|_{z=0} - 2\beta\Omega^2\Big]^{-1}.
\end{equation}
This is a quadratic equation that can be solved for the scale-height $h$, given values for the gas surface density $\Sigma$, the vertical turbulent velocity dispersion $\sigma_z$, the volume densities $\rho_i$ of all disc components, and the galactic rotation curve.

\begin{figure}
\includegraphics[width=\linewidth]{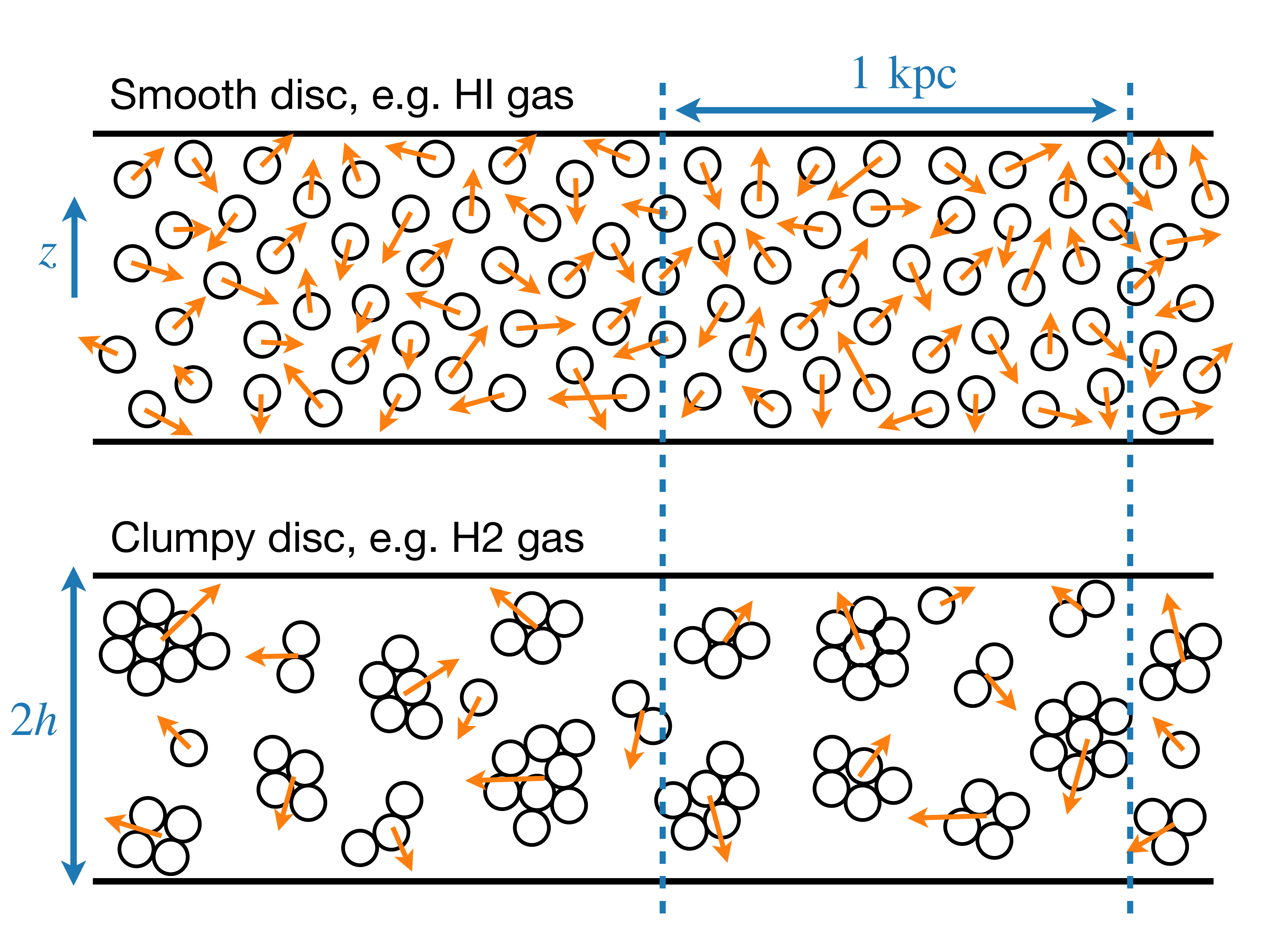}
\caption{Schematic illustrating the difference between a smooth gas disc that can be described by the hydrostatic scale-height using the total kpc-scale molecular gas velocity dispersion (upper half), and a clumpy gas disc that is better described by the hydrostatic prediction using the kpc-scale inter-cloud velocity dispersion (lower half). The orange arrows indicate the velocity vectors of the gas parcels (upper) or gas clumps (lower). The $z$-vector points perpendicular to the disc mid-plane, and $h$ denotes the measured scale-height.}
\label{Fig::clumpy-disc-schematic}
\end{figure}

\subsection{Hydrostatic equilibrium in a clumpy, molecular gas disc} \label{Sec::clumpy-hyd}
The molecular interstellar medium is not smooth, but is organised into cold, dense giant molecular clouds~\citep{Kennicutt+Evans12} of approximate sizes $5$-$200$~pc. Because the clouds are largely confined by self-gravity~\citep[e.g.][]{Sun18} it is unlikely that their internal turbulent velocity dispersions contribute to the effective pressure term in the hydrostatic equation~(\ref{Eqn::hyd-eqm-general}). The observed total vertical velocity dispersion of the molecular gas in a chunk of the disc that encloses a large number of molecular clouds can be written as
\begin{equation} \label{Eqn::kpc-veldisp}
\sigma_{z, {\rm tot}} = \sqrt{\sigma^2_{z, {\rm cl-cl}} + \sigma^2_{z, {\rm int}}},
\end{equation}
where $\sigma_{z, {\rm cl-cl}}$ is the vertical velocity dispersion of the molecular cloud centroids (the \textit{cloud-cloud} velocity dispersion), and $\sigma_{z, {\rm int}}$ is the mean \textit{internal} vertical velocity dispersion of the gas in those clouds. In our Milky Way-like simulation, kpc-scale radial annuli enclose a sufficiently-large number of individual molecular clouds for the value of $\sigma_{\rm cl-cl}$ to be well-defined, corresponding to a minimum of 30 cloud centroids at a degraded spatial resolution of $150$~pc. We describe how each of the velocity dispersions $\sigma_{z, {\rm tot}}$, $\sigma_{z, {\rm cl-cl}}$ and $\sigma_{z, {\rm int}}$ may be measured in Sections~\ref{Sec::kpc-prediction}, \ref{Sec::contour-clouds} and \ref{Sec::single-px-clouds}. As a sanity check, we have also explicitly verified that Equation~(\ref{Eqn::kpc-veldisp}) holds for the three velocity dispersions we measure using our simulation, at all galactocentric radii. This is to be expected, as the values of $\sigma_{z, {\rm cl-cl}}$ and $\sigma_{z, {\rm int}}$ are statistically-independent by construction.

In the following sections we test two different forms of Equation~(\ref{Eqn::sh-v2}) for the molecular disc scale-height of a simulated Milky Way-like galaxy. We first test the case that the molecular gas disc is sufficiently smooth for its scale-height to be calculated using the total kpc-scale molecular gas velocity dispersion $\sigma_{z, {\rm tot}}$, such that
\begin{equation} \label{Eqn::sh-kpc}
h_{\rm tot}^2 = \sigma^2_{z, {\rm tot}} \Big[\frac{2\sqrt{2\pi}G\Sigma_{\rm H_2}}{h_{\rm tot}} + 4\pi G(\rho_{\rm HI}+\rho_*)\Big|_{z=0} - 2\beta\Omega^2\Big]^{-1}.
\end{equation}
This scenario is depicted in the upper half of Figure~\ref{Fig::clumpy-disc-schematic}, where the orange arrows represent the velocity vectors of parcels of gas (equivalent to gas cells in our simulation).

In our second scenario, the internal molecular cloud velocity dispersion $\sigma_{z, {\rm int}}$ is substantially elevated by the locally-enhanced gravitational field of the cloud itself, so that $\sigma_{z, {\rm tot}}$ contains a contribution that does not contribute to the effective pressure supporting the molecular gas disc against the large-scale gravitational field of the galaxy. In this case, Equation~(\ref{Eqn::sh-kpc}) will over-estimate the true molecular disc scale-height, and a better prediction will be given by
\begin{equation} \label{Eqn::sh-inter}
h_{\rm cl-cl}^2 = \sigma^2_{z, {\rm cl-cl}} \Big[\frac{2\sqrt{2\pi}G\Sigma_{\rm H_2}}{h_{\rm cl-cl}} + 4\pi G(\rho_{\rm HI}+\rho_*)\Big|_{z=0} - 2\beta\Omega^2\Big]^{-1},
\end{equation}
where $\sigma_{z, {\rm cl-cl}}$ is measured on kpc scales, and excludes the contributions of turbulent motions on sub-cloud scales. We emphasise that $h_{\rm cl-cl}$ is the vertical scale-height of the cloud ensemble distribution, and not simply the vertical extent of individual clouds. This scenario is shown in the lower half of Figure~\ref{Fig::clumpy-disc-schematic}, where the orange arrows represent the velocity vectors of confined clouds. The lower panel of Figure~\ref{Fig::clumpy-disc-schematic} has the same gas surface density as the upper panel, on kpc-scales. The two cases differ only in their spatial matter distributions, which produces differing gravitational forces between particles. The velocity dispersions in the smooth and clumpy cases are therefore likely substantially different.

By comparing the predictions of Equations~(\ref{Eqn::sh-kpc}) and~(\ref{Eqn::sh-inter}) to the true scale-height $h_{\rm H_2} = \Sigma_{\rm H_2}/\sqrt{2\pi} \rho_{\rm H_2}|_{z=0}$ of the molecular gas disc measured in our simulations, we can determine which value of the observable velocity dispersion ($\sigma_{z, {\rm tot}}$ or $\sigma_{z, {\rm cl-cl}}$) provides the most accurate estimate of the vertical scale-height of the molecular gas.

\begin{figure}
\includegraphics[width=\linewidth]{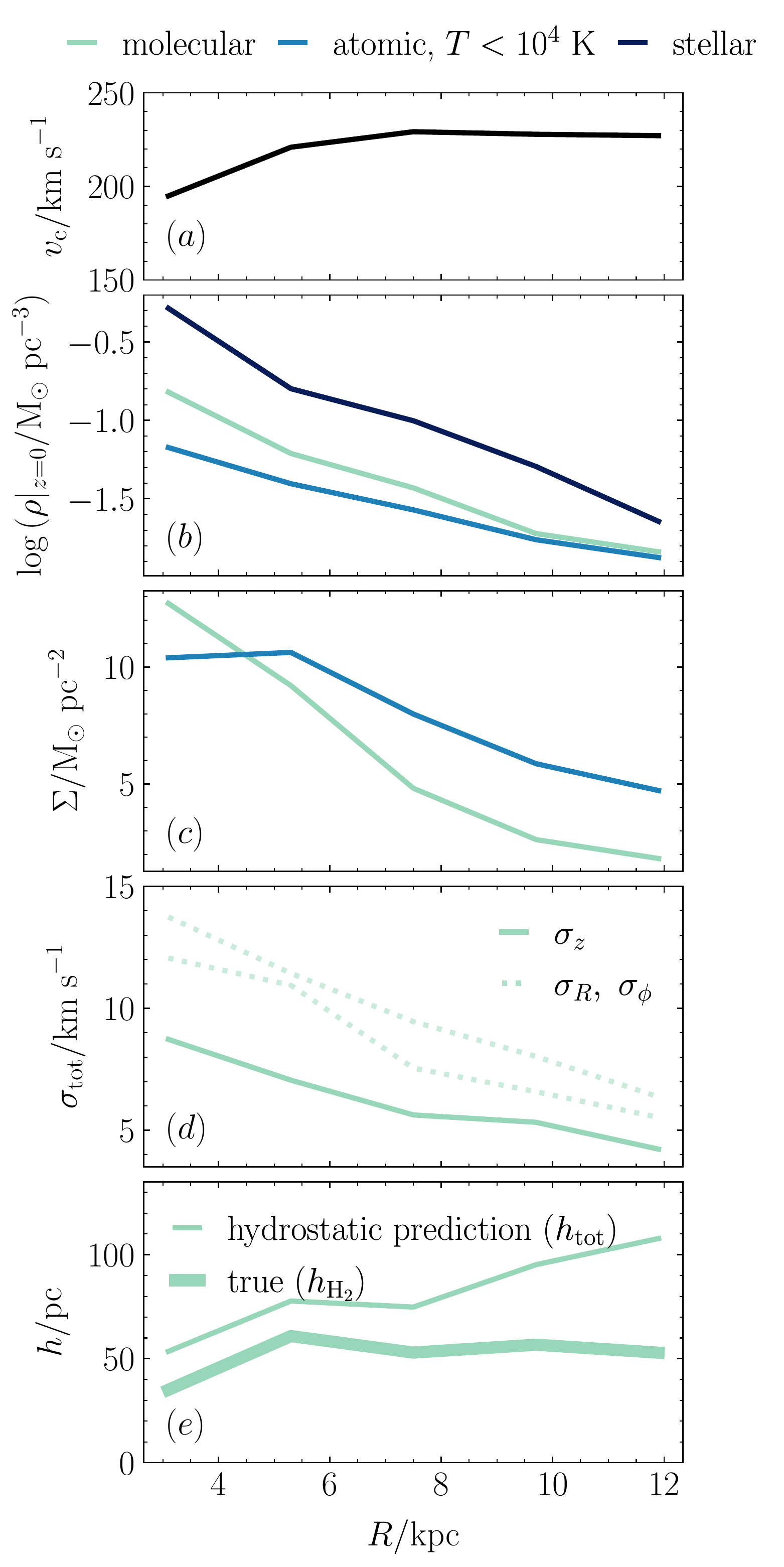}
\caption{The true molecular disc scale-height $h_{\rm H_2} = \Sigma_{\rm H_2}/\sqrt{2\pi}\rho|_{z=0}$ (panel e, thick line) is not consistent with the scale-height $h_{\rm tot}$ predicted by the condition of hydrostatic equilibrium (Equation~\ref{Eqn::sh-kpc}, panel e, thin line) if we assume that the disc is vertically-supported by the total galactic-scale (kpc-scale) molecular gas velocity dispersion $\sigma_{z, {\rm tot}}$ (panel d, solid line). We also demonstrate the slight anisotropy of the total kpc-scale molecular gas velocity dispersion by comparing $\sigma_z$ to the in-plane components $\sigma_R$ and $\sigma_\phi$ (panel d, dotted lines, where $\sigma_R>\sigma_\phi$). Above, we show all of the galactic-scale observables used to compute $h_{\rm tot}$ as a function of the galactocentric radius $R$, including the rotation curve (panel a), the gas and stellar volume densities (panel b), and the gas surface densities (panel c). See Section~\ref{Sec::sh-sims} for futher details.}
\label{Fig::densities-rotcurve}
\end{figure}

\section{Measurement of the molecular gas disc scale-height} \label{Sec::sh-sims}
In Section~\ref{Sec::theory}, we have shown that under the assumption of hydrostatic equilibrium, the scale-height of the molecular gas disc can be derived from measurements of five different observables: the stellar and atomic gas mid-plane volume densities $\rho_*|_{z=0}$ and $\rho_{\rm HI}|_{z=0}$, the galactic rotation curve $v_{\rm c}$, the total kpc-scale molecular gas surface density $\Sigma_{\rm H_2}$ and the molecular gas velocity dispersion $\sigma_{z, {\rm H_2}}$, perpendicular to the galactic plane. We note that $\Sigma_{\rm H_2}$ is relevant for the galactic-scale hydrostatic equilibrium calculation because all the molecular gas mass in a kpc-scale chunk of the disc contributes to the large-scale gravitational potential. The values of $\rho_*|_{z=0}$, $\rho_{\rm HI}|_{z=0}$, $v_{\rm c}$ and $\Sigma_{\rm H_2}$ are independent of the cloud-scale gas distribution. By contrast, for a clumpy medium, there are two possible choices of $\sigma_{z, {\rm H_2}}$. If the gravitational field of the molecular gas is strongly enhanced by the local self-gravity within molecular clouds, the cloud-cloud velocity dispersion $\sigma_{z, {\rm cl-cl}}$ (Equation~\ref{Eqn::sh-inter}) will likely provide a more accurate estimate than the the total molecular gas velocity dispersion $\sigma_{z, {\rm tot}}$ (Equation~\ref{Eqn::sh-kpc}).

\subsection{Hydrostatic scale-height predicted using $\sigma_{z, {\rm tot}}$} \label{Sec::kpc-prediction}
In Figure~\ref{Fig::densities-rotcurve}, we show radial profiles of the observables in Equation~(\ref{Eqn::sh-kpc}), and the resulting hydrostatic scale-height, computed within annuli of width $2.2$~kpc encircling the galactic centre between galactocentric radii of $2$ and $13$~kpc:
\begin{enumerate}
	\item \textit{Panel (a):} The circular velocity $v_{\rm c}$, derived from the velocity vectors of all gas cells in the simulation. It is approximately flat, so the $-2\beta \Omega^2$ term in Equation~(\ref{Eqn::sh-v2}) is small relative to the other terms.
	\item \textit{Panel (b):} The atomic and stellar mid-plane volume densities $\rho_{\rm HI}|_{z=0}$ and $\rho_*|_{z=0}$ (blue and dark blue lines), given by the maximum values of the vertical profiles of the volume densities at each galactocentric radius (see Figure~\ref{Fig::z-rho-dstbn}). The annulus-averaged molecular mid-plane volume density is also shown, for reference (aqua line).
	\item \textit{Panel (c):} The molecular gas surface density $\Sigma_{\rm H_2}$ (aqua line), given by the total molecular gas mass in each radial bin, divided by its area. The atomic gas surface density is also shown, for reference (blue line).
	\item \textit{Panel (d):} The total vertical kpc-scale molecular gas velocity dispersion $\sigma_{z, {\rm tot}}$. We subtract a two-dimensional projection of the mass-weighted mean vertical velocity at a map resolution of $1$~kpc from the vertical velocities of all gas cells in the simulation, then compute the dispersion of the resulting velocities in each radial bin.\footnote{For completeness here and in later calculations, we include the thermal contribution to the velocity dispersion by adding the sound speed in the molecular gas ($c_s \sim 0.2~{\rm km~s}^{-1}$) to the supersonic turbulent component in quadrature. This makes a negligible contribution to the total velocity dispersion.} We also show the radial profiles of the in-plane components of the velocity dispersion for reference. We see that the velocity dispersion is substantially anisotropic on kpc-scales.
	\item Finally, the thin line in panel (e) of Figure~\ref{Fig::densities-rotcurve} is computed from all of the above observables using Equation~(\ref{Eqn::sh-kpc}). It lies substantially above the true molecular disc scale-height (thick line, $h_{\rm H_2} = \Sigma_{\rm H_2}/\sqrt{2\pi}\rho_{\rm H_2}|_{z=0}$) derived from the full three-dimensional distribution of gas volume densities.
\end{enumerate}
Figure~\ref{Fig::densities-rotcurve} therefore demonstrates that using the total molecular gas velocity dispersion in the hydrostatic equation~(\ref{Eqn::sh-kpc}) produces an incorrect prediction for the molecular disc scale-height. At small galactocentric radii $R \sim 2$~kpc, Equation~(\ref{Eqn::sh-inter}) over-estimates the molecular scale-height by $25$~pc. As $R$ increases, the hydrostatic prediction displays `flaring' behaviour and increases up to $110$~pc, while the true molecular disc scale-height remains flat at $50$~pc.

In the following sub-sections, we derive the alternative form of the molecular gas disc scale-height predicted by Equation~(\ref{Eqn::sh-inter}), using the velocity dispersion $\sigma_{z, {\rm cl-cl}}$ between molecular cloud centroids, rather than the total molecular gas velocity dispersion $\sigma_{z, {\rm tot}}$. We determine whether the former is a more accurate estimate of the gas motion counteracting the large-scale gravitational field in the galaxy. We therefore determine whether Equation~(\ref{Eqn::sh-inter}) provides a better prediction for the molecular disc scale-height than does Equation~(\ref{Eqn::sh-kpc}).

\begin{figure}
\includegraphics[width=\linewidth]{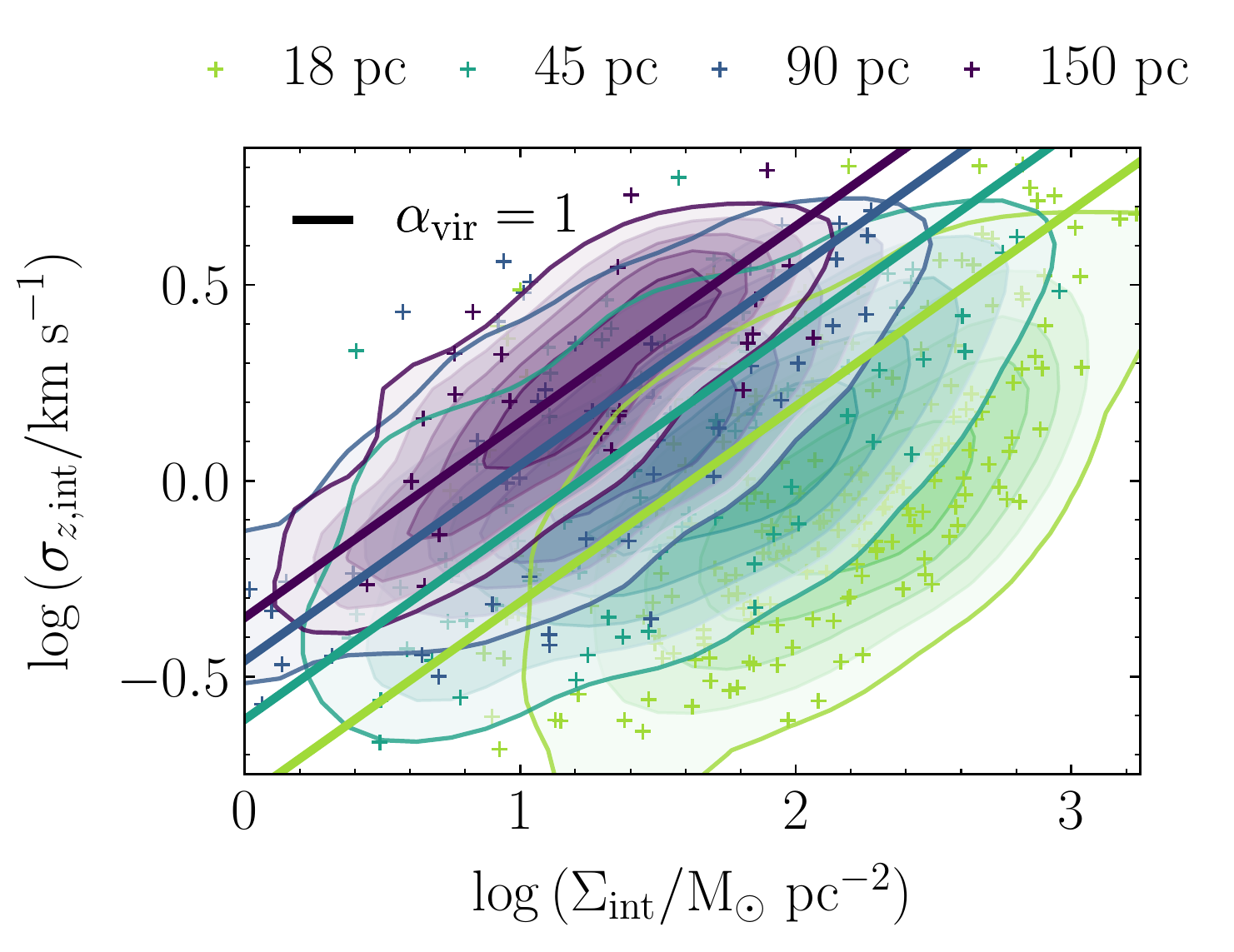}
\caption{Internal line-of-sight velocity dispersion $\sigma_{z, {\rm int}}$ as a function of the molecular cloud surface density $\Sigma_{\rm int}$ for the molecular clouds identified via the single-pixel method at map resolutions of $150$~pc (purple), $90$~pc (blue), $45$~pc (turquoise) and $18$~pc (green). The normalised parameter-space density of the clouds is enclosed by the contours and one twentieth of the identified clouds are additionally shown as crosses. A virial parameter of $\alpha_{\rm vir}=1$ for spherical beam-filling clouds at each resolution is given by the solid lines.}
\label{Fig::bw-diagnostics}
\end{figure}

\begin{figure*}
\includegraphics[width=\linewidth]{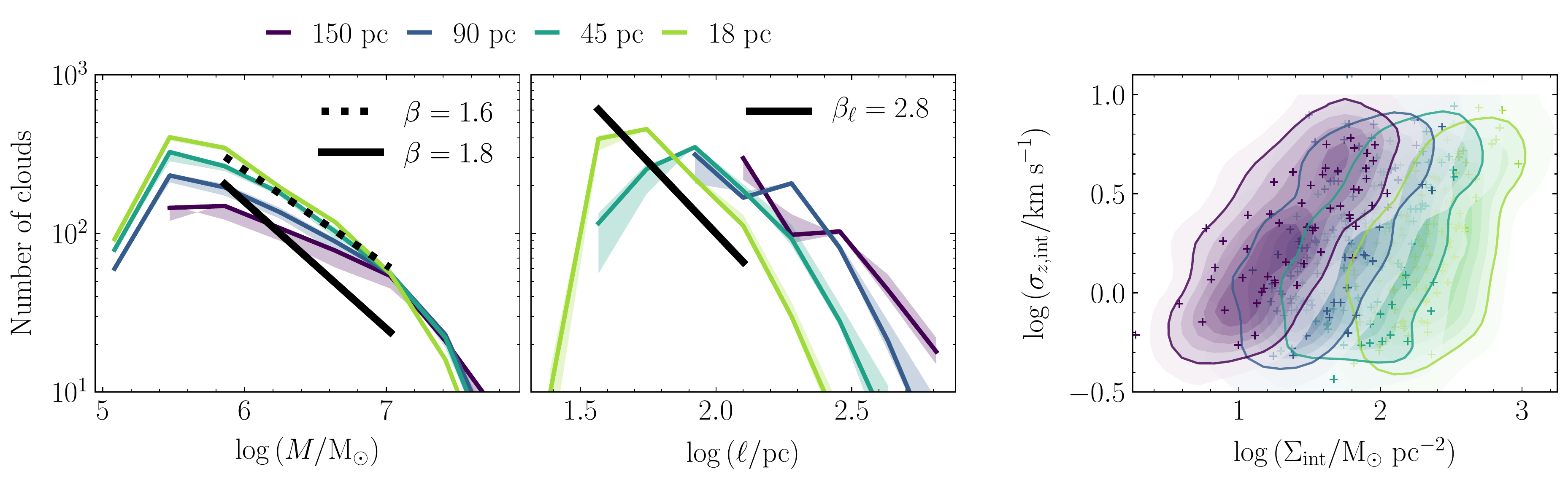}
\caption{{\it Left panel:} Mass distribution of the molecular clouds identified via the contour-cloud method at map resolutions of $150$~pc (purple), $90$~pc (blue), $45$~pc (turquoise) and $18$~pc (green). The solid black and dashed lines denote the range of power-law slopes for the observed cloud mass distribution in the Milky Way, given by $\dd N/\dd M\propto M^{-\beta}$ with $\beta \in [1.6, 1.8]$. {\it Centre panel:} Size distribution of the clouds. The black line denotes the power-law slope of the observed cloud size distribution in the Milky Way, given by $\dd N/\dd \ell \propto \ell^{-\beta_\ell}$ with $\beta_\ell \sim 2.8$. The shaded regions in both panels show the variation in the mass/size spectra with the cloud-identification threshold $\Sigma_{\rm thresh}$ over the range from $3$ to $12~{\rm M}_\odot~{\rm pc}^{-2}$. Solid lines give the values corresponding to $\Sigma_{\rm thresh} = 10~{\rm M}_\odot~{\rm pc}^{-2}$. {\it Right panel:} Internal line-of-sight molecular gas velocity dispersion $\sigma_{z, {\rm int}}$ as a function of cloud surface density $\Sigma_{\rm int}$ at each map resolution. The parameter-space density of the clouds is enclosed by the contours and one tenth of the identified clouds are additionally shown as crosses.}
\label{Fig::cprops-diagnostics}
\end{figure*}

\subsection{Molecular cloud samples} \label{Sec::cloud-samples}
We test two different methods for separating molecular clouds from the surrounding interstellar medium, both of which are used in the observational literature. Both methods require two-dimensional maps of the molecular gas surface density $\Sigma_{\rm H_2}$, which we create by post-processing our simulation output with a chemistry and radiative-transfer model, to produce realistic molecular gas abundances (see  Appendix~\ref{App::Sigma-H2-maps} for further details). We compute $\Sigma_{\rm H_2}$ maps at map resolutions of $\epsilon = 150$~pc and $90$~pc, to match the latest observations of molecular gas across the sample of $70$ nearby galaxies from the PHANGS collaboration~\citep[see][]{2020ApJ...901L...8S,2021arXiv210407739L}. We additionally compute maps at resolutions of $\epsilon = 45$~pc and $18$~pc to probe the smaller scales that may be observable for external galaxies in the near future.

\subsubsection{Single-pixel clouds}
Our first method for cloud identification selects `single-pixel clouds' or `molecular gas sight-lines', as advocated by~\cite{2016ApJ...831...16L,Sun18}. Each pixel with $\Sigma_{\rm H_2}>\Sigma_{\rm thresh}$ is counted as a separate molecular cloud. We test three values $\Sigma_{\rm thresh} = 3, 10$~and~$50~{\rm M_\odot}~{\rm pc}^{-2}$ of the threshold for cloud identification, to demonstrate that our results are not significantly altered by the observational sensitivity limit. This method has the advantage of evenly-sampling the molecular gas across the area of the galactic disc, rather than assigning fewer measurements to large, contiguous areas of CO emission. Such assignments will vary according to the specific clump-finding procedure used.

\subsubsection{Contour clouds}
Our second method for cloud identification selects `contour-clouds' identified in position-position-velocity (PPV) space. We use the same maps of $\Sigma$ as for the single-pixel clouds, but select contiguous regions of CO-bright molecular hydrogen from the trunk of the dendrogram produced by applying the {\sc Astrodendro} package~\citep{2008ApJ...679.1338R} with a minimum column density of $\Sigma_{\rm H_2} > \Sigma_{\rm thresh}$. In order to determine if there are multiple, separated clouds along the same line of sight, we divide the simulated gas cells into velocity channels according to their line-of-sight velocities. We use a channel width of $2.5$~km/s, corresponding to the resolution used for cloud identification in PPV space for the PHANGS sample~\citep{2021MNRAS.502.1218R}. We compute the molecular gas mass in each channel and identify secondary peaks along the $v_z$ axis that have a prominence of at least half the height of the primary (tallest) peak. In the case that there are multiple peaks within a single two-dimensional contour, we divide the masked gas cells into separate clouds at the value of $v_z$ associated with the local minimum between the peaks. We find that the fraction of two-dimensional clouds with multiple peaks along the $v_z$ axis varies between zero and seven~per~cent at all map resolutions $\epsilon$ for a peak prominence of $0.5$. At a lower peak prominence threshold of $0.25$, this fraction rises to 12~per~cent at the smallest galactocentric radii $R<5$~kpc, and retains a similar value at $R>5$~kpc. Accordingly, our choice of peak prominence threshold has a negligible effect on the distribution of cloud centroid velocities and internal cloud velocity dispersions at all map resolutions.

\subsubsection{Observational checks of simulated clouds}
In Figures~\ref{Fig::bw-diagnostics} and~\ref{Fig::cprops-diagnostics}, we check the key observable diagnostics for the single-pixel and contour cloud populations at each map resolution. For the single-pixel clouds we show the relationship between the cloud surface density $\Sigma_{\rm int}$ and the internal line-of-sight velocity dispersion $\sigma_{z, {\rm int}}$. We take a lower threshold of $3 \times 10^4~{\rm M}_\odot$ on the cloud mass to ensure that each cloud includes enough gas cells ($>30$) for us to reliably measure $\sigma_{z, {\rm int}}$. By comparison to Figure 4 of~\cite{Sun18}, we see that our single-pixel cloud populations at map resolutions of $150$ and $90$~pc resolution are comparable to observed molecular gas sightlines at similar resolutions in external galaxies. That is, our clouds follow a line of roughly-constant virial parameter, with a significant fraction of both bound ($\alpha_{\rm vir}<1$) and unbound ($\alpha_{\rm vir}>1$) clouds. At higher map resolutions of $45$ and $18$~pc, the fraction of bound clouds increases as we are able to resolve the true sizes and surface densities of the smaller molecular regions in the simulation.

For the contour-clouds we show the mass and size distributions (left and central panels of Figure~\ref{Fig::cprops-diagnostics}), as well as the size-linewidth relation (right-hand panel), at each map resolution. The typical diameter of a gravitationally-bound molecular region in the simulation is $30$~pc, and the typical region separation is $\sim 100$~pc. This means that the largest contour-clouds at map resolutions of $18$ and $45$~pc represent well-resolved molecular regions. The slopes of their mass and size spectra adhere well to the respective values of $\dd N/\dd M \propto M^{-\beta}, \beta \in [1.6, 1.8]$ and $\dd N/\dd \ell \propto \ell^{-\beta_\ell}, \beta = 2.8$ observed at high spatial resolution within the Milky Way~\citep{Solomon87,Williams&McKee97,Heyer+09,Roman-Duval+10,Miville-Deschenes17,Colombo+19}. As the map resolution is decreased through 90~pc to 150~pc, the mean region size is over-estimated to an increasing extent, as unresolved molecular regions are `smeared out' over an increasingly-large beam. The estimates of cloud mass and velocity dispersion remain relatively constant, because the separation between molecular regions remains well-resolved (one cloud per beam on average, even at $\epsilon = 150$~pc). These effects cause an increase in the average cloud size relative to the cloud mass (compare left-hand and centre panels) and to the velocity dispersion (see right-hand panel).

We note that at the highest map resolution of $18$~pc, the majority of the identified contour-clouds are sub-virial, while sub-virial clouds are very rarely found in the observational literature. This is likely due to a combination of factors. Firstly, we have used a definition of the virial parameter (Equation 6 in~\citealt{Sun18}) that assumes spherical clouds. This may be a particularly poor approximation of the shapes of the highest-resolution contour clouds, which are likely elongated~\citep[see e.g.][]{2020MNRAS.498..385J}. The error introduced by this assumption is of the order $[{\rm pixel~size}^2 \times 2({\rm scale~height})]^{1/3}/{\rm pixel~size} \sim (2\times50~{\rm pc}/{\rm 18~pc})^{1/3} \sim 0.26~{\rm dex}$. Secondly, clouds identified at the highest resolution are most strongly affected by the unresolved turbulence below the native resolution of the simulation ($\sim 6$~pc in our case). The error introduced by unresolved turbulence, assuming a typical Milky Way-like size-linewidth relation~\citep[e.g.][]{Colombo+19} is of the order $1-(5~{\rm km~s}^{-1})/[(5~{\rm km~s}^{-1})^2+(2.5~{\rm km~s}^{-1})^2]^{1/2} \sim 0.1~{\rm dex}$. Together these effects may account for a suppression of the virial parameter of order $\sim 0.4~{\rm dex}$, accounting for the sub-virial objects.

\begin{figure*}
\includegraphics[width=\linewidth]{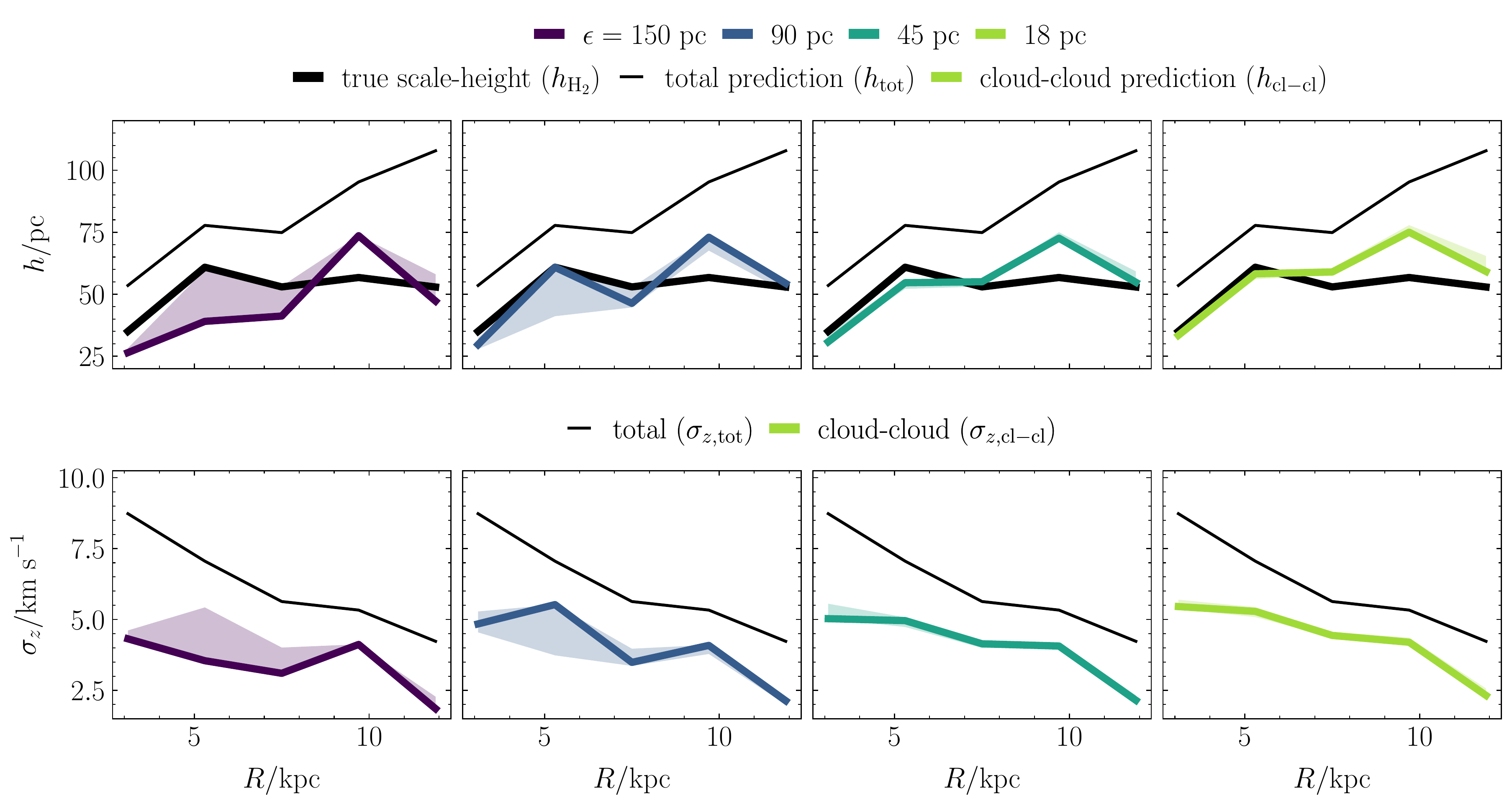}
\caption{The upper panels show that the hydrostatic scale-height $h_{\rm cl-cl}$ predicted using the cloud-cloud velocity dispersion (thick coloured lines) is a better match for the true simulated molecular disc scale-height $h_{\rm H_2}$ (thick black lines) than is the prediction $h_{\rm tot}$ using the total kpc-scale molecular gas velocity dispersion (thin black lines). The lower panels show the corresponding values of the turbulent velocity dispersion, $\sigma_{z, {\rm cl-cl}}$ and $\sigma_{z, {\rm tot}}$. The molecular clouds are identified via the contour-cloud method (see Section~\ref{Sec::cloud-samples}). The fiducial value of the surface density threshold for cloud identification is $\Sigma_{\rm thresh} = 10~{\rm M}_\odot~{\rm pc}^{-2}$, and the shaded regions show the values of the scale-height and velocity dispersion obtained for clouds identified at surface density thresholds from $3$ to $50~{\rm M}_\odot~{\rm pc}^{-2}$.}
\label{Fig::2D-result-cprops}
\end{figure*}

\subsection{The molecular disc scale-height using contour-clouds} \label{Sec::contour-clouds}
In Figure~\ref{Fig::2D-result-cprops}, we compare the true scale-height of the molecular gas disc to (1) the hydrostatic prediction using the total molecular gas velocity dispersion from Section~\ref{Sec::kpc-prediction} and (2) the hydrostatic prediction using the cloud-cloud velocity dispersion of the clouds identified in Section~\ref{Sec::cloud-samples}.

The total molecular gas velocity dispersion $\sigma_{z, {\rm tot}}$ (thin black lines, lower panels) incorporates turbulence at scales ranging from $1$~kpc down to the resolution limit of our simulation at $6$~pc, and ranges from $\sim 9~{\rm km~s}^{-1}$ at small galactocentric radii down to $\sim 4~{\rm km~s}^{-1}$ at large radii. As first shown in Figure~\ref{Fig::densities-rotcurve}, the corresponding prediction $h_{\rm tot}$ for the molecular disc scale-height using Equation~\ref{Eqn::sh-kpc} (thin black lines, upper panels of Figure~\ref{Fig::2D-result-cprops}) exceeds the true scale-height $h_{\rm H_2}$ by around $15$~pc at small galactocentric radii and by $60$~kpc at the edge of the galactic disc. It has a radially-averaged value of $80$~pc, around $30$~pc larger than the true radially-averaged scale-height of $50$~pc.

The cloud-cloud velocity dispersion $\sigma_{z, {\rm cl-cl}}$ (thick coloured lines, lower panels) is calculated between the centroids of molecular clouds identified via the contour-cloud method (see Section~\ref{Sec::cloud-samples}) at each of the map resolutions $\epsilon = 18, 45, 90$ and $150$~pc. The centroid velocity of each identified cloud is computed as a molecular mass-weighted average over the vertical velocities of the gas cells it contains. The kpc-averaged velocity field of the molecular gas is subtracted from these cloud centroid velocities, and their cloud mass-weighted velocity dispersion is calculated within each radial annulus. The cloud-cloud velocity dispersion therefore incorporates turbulence at scales from $1$~kpc down to the cloud scale, while excluding contributions at smaller, sub-cloud scales.

At all map resolutions, the corresponding scale-height $h_{\rm cl-cl}$ computed via Equation~\ref{Eqn::sh-inter} (thick coloured lines, upper panels) is a better predictor of the true molecular gas disc scale-height than is $h_{\rm tot}$. In particular, $h_{\rm cl-cl}$ captures the near-constant value of the true scale-height at galactocentric radii larger than $5$~kpc, exhibiting none of the flaring behaviour seen for $h_{\rm tot}$. This difference in behaviour indicates that a significant contribution to the total kpc-scale velocity dispersion $\sigma_{z, {\rm tot}}$ comes from sub-cloud scales, but that turbulence at these small scales does not contribute meaningfully to the hydrostatic support of the molecular gas disc.

\begin{figure*}
\includegraphics[width=\linewidth]{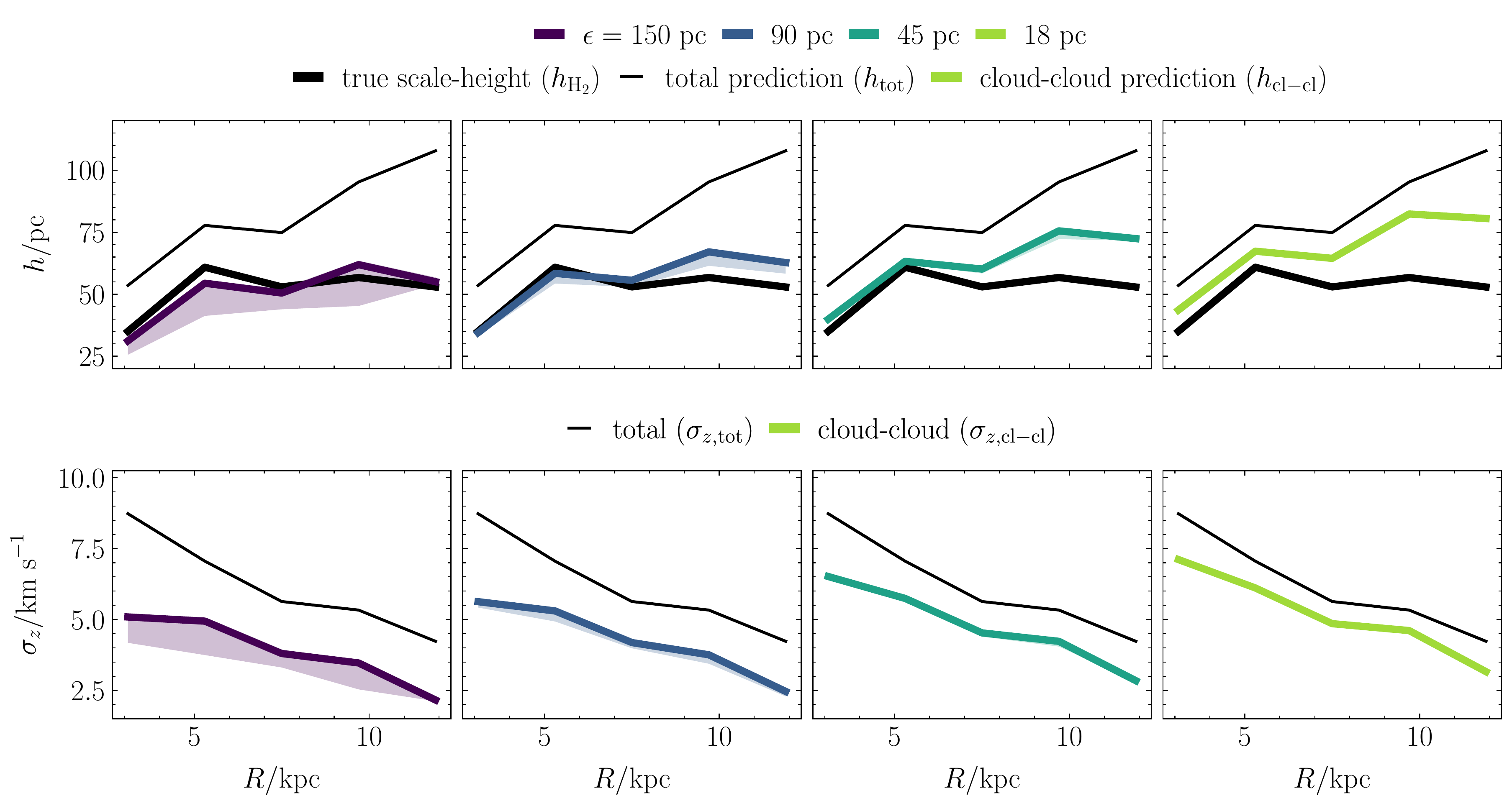}
\caption{Similar to Figure~\ref{Fig::2D-result-bw}, but for molecular clouds identified via the single-pixel method.}
\label{Fig::2D-result-bw}
\end{figure*}

We note that the value of $\sigma_{z, {\rm cl-cl}}$ calculated using contour clouds at a map resolution of $150$~pc provides a worse (lowered) estimate of the disc scale-height than the values at higher resolutions. This is because the average separation of $\sim 100$~pc between the centroids of the self-gravitating clouds in our simulation is unresolved at $150$~pc resolution, effectively grouping some clouds together and so removing their contribution to $\sigma_{z, {\rm cl-cl}}$.

\subsection{The molecular disc scale-height using single-pixel clouds} \label{Sec::single-px-clouds}
In Figure~\ref{Fig::2D-result-bw}, we repeat the analysis described in Section~\ref{Sec::contour-clouds} for molecular clouds identified via the single-pixel method. Similar to the case of clouds identified using the contour-cloud method, the scale-height derived using the cloud-cloud velocity dispersion is closer to the true molecular disc scale-height $h_{\rm H_2}$ than is the scale-height $h_{\rm tot}$ derived from the total molecular gas velocity dispersion. However, in the case of single-pixel clouds, close agreement between the true scale-height and the cloud-cloud hydrostatic prediction is only obtained at the lowest map resolutions, $\epsilon = 150$~pc and $90$~pc. As the map resolution is increased from $\epsilon = 90$~pc through $\epsilon = 45$ and $18$~pc, the cloud-cloud velocity dispersion increases monotonically. At $\epsilon = 18$~pc, it is close to the total velocity dispersion $\sigma_{z, {\rm tot}}$.

This trend can be explained in terms of the properties of gravitationally-bound molecular regions in the simulation, which have a maximum size of $\sim 120$~pc and a minimum separation of $\sim 100$~pc. At map resolutions of $90$ and $150$~pc, the single-pixel method assigns approximately one bound region per pixel. By contrast, at the highest map resolution ($\epsilon = 18$~pc), the single-pixel method divides the bound regions into multiple pixels, and the velocity dispersion between these pixels is mistakenly counted towards the cloud-cloud velocity dispersion. The contour-cloud approach is more stable with changing map resolution because it defines clouds based on their isodensity contours.

\section{Discussion} \label{Sec::discussion}
The results presented in Section~\ref{Sec::sh-sims} imply that the molecular gas disc is in an approximate state of hydrostatic equilibrium, in which the velocity dispersion between individual clouds (the cloud-cloud velocity dispersion) balances the gravitational force acting on these clouds. We emphasise that the common approach of using the total molecular gas velocity dispersion could lead to over-estimated scale-heights because it includes the velocity dispersion internal to the cloud, $\sigma_{z, {\rm int}}$. While $\sigma_{z, {\rm int}}$ reflects the internal dynamics of molecular clouds, it does not directly contribute to supporting the molecular gas disc on large scales. Here we compare our findings to observations of the molecular disc scale-height in the Milky Way and external galaxies, and to previous investigations of hydrostatic equilibrium in numerically-simulated Milky Way-like disc galaxies. We state the limitations of our simulation and discuss their possible effects on our results.

\subsection{Comparison to observations} \label{Sec::obs-comp}
\subsubsection{Observations of the Milky Way} \label{Sec::obs-MW}
The most reliable measurements of the molecular disc scale-height in the Milky Way can be made between galactocentric radii of $R\sim 2.5$~kpc and the solar radius at $R \sim 8$~kpc~\citep{2015ARA&A..53..583H}. Of these, the study of~\cite{1994ApJ...433..687M} provides the most accurate distances to the molecular clouds used, obtaining an average molecular disc full-width half maximum (FWHM) that varies between $90$ and $120$~pc. This corresponds to a Gaussian scale-height that varies between $38$ and $50$~pc, in very good agreement with the molecular disc scale-height measured for our simulation, $h_{\rm H_2}$. Outside the solar radius, measurements are limited to narrow ranges in Galactic longitude~\citep{1987ApJ...315..122G,1988ApJ...327..139C,1991PhDT.........5D} and so vary widely, but tentatively display uniformly-higher values than within the solar circle. We emphasise that this `flaring' of the molecular gas disc is \textit{not} reproduced in our simulation. If it exists, it is likely caused by physics not present in an isolated simulation (e.g.~a past merger with another galaxy). Although the overestimate of the molecular disc scale-height caused by assuming hydrostatic equilibrium supported by the kpc-scale molecular gas velocity dispersion \textit{does} display disc flaring, we have shown that this is an artefact caused by the inclusion of sub cloud-scale, self gravity-driven turbulence.

\subsubsection{Observations of external galaxies} \label{Sec::obs-ext-gal}
\begin{figure}
\includegraphics[width=\linewidth]{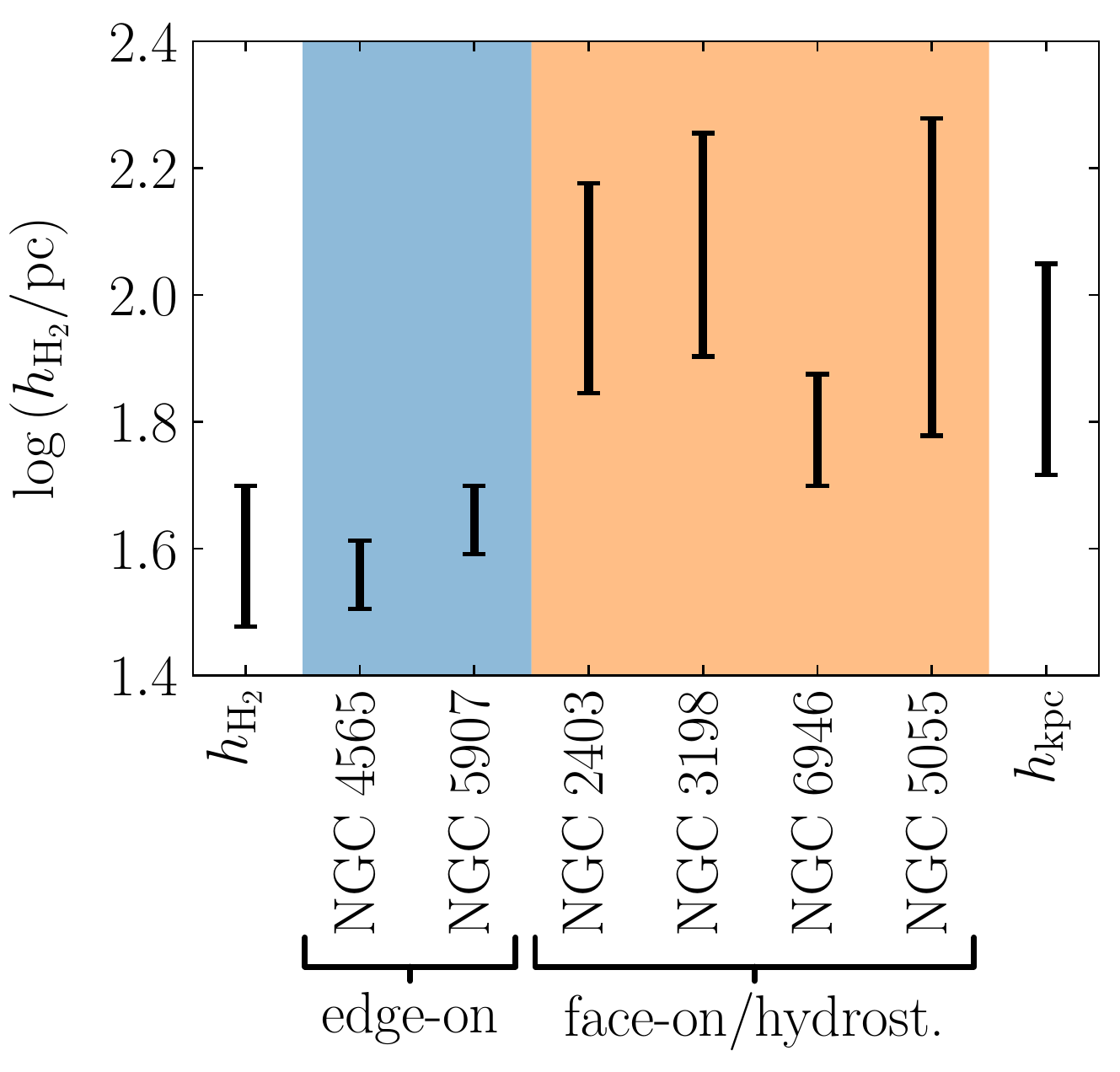}
\caption{The ranges of observed molecular disc scale-heights in external galaxies with edge-on viewing angles (shaded blue) agree well with the molecular disc scale-height $h_{\rm H_2}$ of our simulated galaxy. The ranges of hydrostatic molecular disc scale-heights derived using observations of the total molecular gas velocity dispersion in face-on galaxies (shaded orange) are consistent with our hydrostatic scale-height using the total kpc-scale velocity dispersion, $h_{\rm tot}$. On average, they are $60$~pc larger than the observed values for edge-on galaxies.}
\label{Fig::obs-boxplot}
\end{figure}

The molecular disc scale-height has been observed in external edge-on disc galaxies by~\cite{2011AJ....141...48Y,2014AJ....148..127Y}. This quantity has also be inferred based on the hydrostatic assumption in face-on and inclined disc galaxies by~\cite{2019A&A...622A..64B}, and in luminous and ultra-luminous infrared galaxies by~\cite{2019ApJ...882....5W}. The sample of observed gas discs that most closely resemble our Milky Way-like simulation consists of NGC 4565 and NGC 5907 from~\cite{2014AJ....148..127Y}, and NGC 2403, NGC 3198, NGC 6946 and NGC 5055 from~\cite{2019A&A...622A..64B}. These galaxies have molecular and atomic gas surface densities in the ranges $\Sigma_{\rm H_2} \in [2,20]~{\rm M}_\odot~{\rm pc}^{-2}$ and $\Sigma_{\rm HI} \in [4,16]~{\rm M}_\odot~{\rm pc}^{-2}$, respectively, and kpc-scale atomic gas velocity dispersions in the range of $\sigma_{z, {\rm tot}} \in [10,20]~{\rm km~s}^{-1}$ for galactocentric radii $R>2$~kpc, close to the values we have reported in Figure~\ref{Fig::densities-rotcurve}. In Figure~\ref{Fig::obs-boxplot} we show the ranges of molecular gas disc scale-heights reported for these galaxies at galactocentric radii of $R>2$~kpc, in comparison to the range of the true molecular disc scale-height $h_{\rm H_2}$ and the kpc-scale hydrostatic molecular disc scale-height $h_{\rm tot}$ for our simulated galaxy. Edge-on observations (shaded blue) allow for the direct determination of the scale-height from the intensity profile of observed CO emission. These values agree most-closely with the true scale-height of our simulated galaxy. Face-on observations (shaded orange) require the assumption of hydrostatic equilibrium, and rely on a molecular gas velocity dispersion measured on kpc scales, due to limited data resolution. As expected, these estimates are consistent with our scale-height $h_{\rm tot}$, inferred using the simulated total kpc-scale molecular gas velocity dispersion in the hydrostatic equation. Although the sample is small, it appears that hydrostatic scale-heights that rely on kpc-scale velocity dispersion measurements tend to be larger than the scale-heights that are directly observed, and we suggest that this discrepancy can be attributed to the inclusion of the sub-cloud component of the molecular gas velocity dispersion in the hydrostatic equation. We predict that if the cloud-cloud velocity dispersion $\sigma_{z, {\rm cl-cl}}$ were used in place of the total molecular gas velocity dispersion for the galaxies NGC 2403, 3198, 6946 and 5055 ($\sigma_{z, {\rm cl-cl}}$ has already been calculated in the galaxies NGC 2404 and 5055 by~\citealt{2011MNRAS.410.1409W}), the derived scale-height would be smaller in value, and closer to the edge-on scale-heights observed for the galaxies NGC 4565 and 5907. That is, given higher-resolution CO observations resolving individual molecular clouds, the cloud-cloud velocity dispersion would provide a more accurate determination of the true scale-height of the molecular gas disc than the total velocity dispersion.

\begin{figure}
\includegraphics[width=\linewidth]{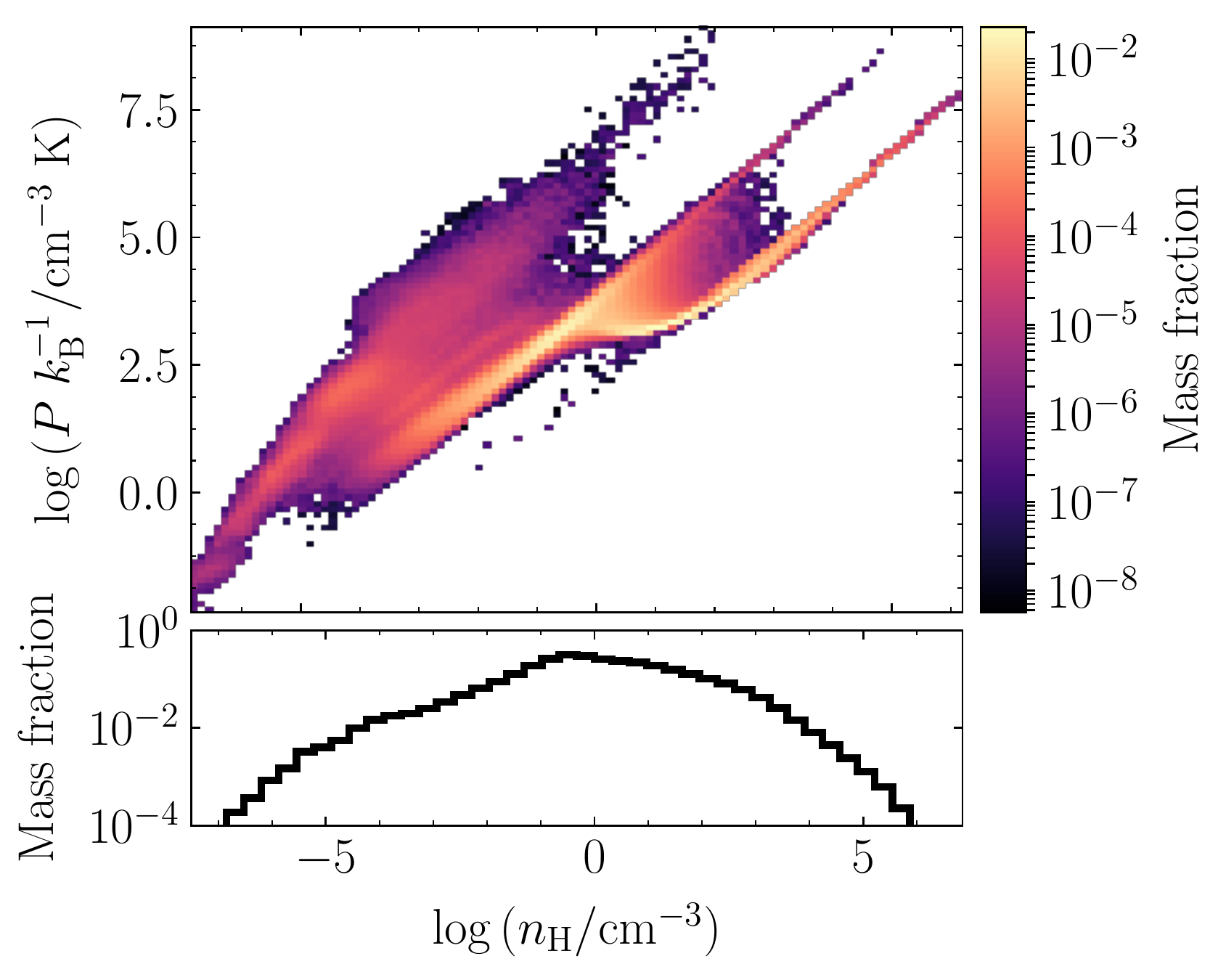}
\caption{Distribution of gas in the density-pressure plane, for the entire simulated disc galaxy. The mass-weighted density histogram is given in the lower panel.}
\label{Fig::phases}
\end{figure}

\subsection{Comparison to numerical simulations}
The study of~\cite{2009ApJ...693.1346K} examines the state of vertical hydrostatic equilibrium in kpc-scale simulations containing a self-gravitating, multi-phase interstellar medium. The gas in each simulation has a Milky Way-like column density and is supported by turbulence driven at large scales by HII region feedback. Similar studies are conducted by~\cite{2011ApJ...743...25K,2013ApJ...776....1K,KimCG&Ostriker15b} (see also~\citealt{IffrigHennebelle15}), but focus in particular on the diffuse gas reservoir, in accordance with the theory of~\cite{Ostriker+10}. The work of~\cite{2009ApJ...693.1346K} is therefore most directly-comparable to our results.

\cite{2009ApJ...693.1346K} compare the true scale-height and mid-plane pressure of the gas to the hydrostatic predictions for each quantity, using the mass-weighted kpc-scale velocity dispersion of the total gas reservoir (diffuse plus dense gas).\footnote{We have used the molecular mass-weighted velocity dispersion rather than the total gas mass-weighted value, but given that high molecular fractions are associated with the densest gas in our simulation, we expect the values to be comparable.} They find good agreement between the true and hydrostatic values, in apparent contradiction with our results. However, this tension can be explained by noting the much larger fraction of gas in our simulations that is contained at high densities $n_{\rm H}>100~{\rm cm}^{-3}$. This difference can be seen by comparing our Figure~\ref{Fig::phases} to Figure 4 of~\cite{2009ApJ...693.1346K}. The larger quantity of dense (self-gravitating, molecular) gas in our simulations indicates a larger contribution made by this gas to the mass-weighted kpc scale turbulent velocity dispersion $\sigma_{z, {\rm tot}}$. Certainly, our values of $\sigma_{z, {\rm kpc}} \in [5,10]~{\rm km~s}^{-1}$ are substantially higher than the values $\sigma_{z, {\rm tot}} \in [2,5]~{\rm km~s}^{-1}$ obtained by~\cite{2009ApJ...693.1346K}. Three possible origins for the increased fraction of dense gas in our simulations relative to this work are as follows:
\begin{enumerate}
\item Galactic-dynamical processes are included in our simulations. In particular, we include large-scale shearing motions due to galactic rotation, as well as flocculent spiral arm structures created by large, hot, ionised regions blown in the interstellar medium by supernova feedback. Along these spiral arms and at the edges of these bubbles, gas is compressed into molecular clouds.
\item We compute the gravitational accelerations of individual dark matter and stellar particles (known as a `live gravitational potential’), rather than using a smooth analytic gravitational potential. This approach allows for increased gravity-induced clumping of the simulated dark matter and baryons.
\item A significant fraction of the molecular clouds in our simulations result from mergers of multiple smaller clouds, which aggregate mass into larger, denser structures.
\end{enumerate}
These three differences may all feasibly contribute to a larger reservoir of dense/molecular gas in our simulations, driving up the contribution to the velocity dispersion that is provided by the locally-enhanced gravitational field inside molecular clouds.

\subsection{Caveats}
The primary limitation associated with our galaxy model, with the potential to affect the presented results, is the exclusion of any physics associated with magnetic fields or cosmic rays. A magnetic field permeating the interstellar medium provides an additional vertical pressure term in the equation of hydrostatic equilibrium~\citep[e.g.][]{2007ApJ...663..183P,KimCG&Ostriker15b}. In the ambient cloud-cloud medium, the magnetic pressure term is of the same order as the turbulent pressure~\citep{2003ApJ...586.1067H,2005ApJ...624..773H,2007ApJ...663..183P}, but inside dense molecular gas it is generally observed as sub-dominant to the turbulent pressure~\citep{Crutcher12}. The introduction of such a magnetic pressure may therefore simultaneously increase the true scale-height of the simulated gas disc as well as the hydrostatic prediction using the cloud-cloud velocity dispersion, while having a smaller relative effect on the hydrostatic prediction using the full kpc-scale molecular gas velocity dispersion. However, it should not alter our primary conclusion that the internal molecular cloud velocity dispersion should be excluded for an accurate prediction of the gas disc scale-height. A comparison of our results to the true and hydrostatic molecular disc scale-heights in a full magnetohydrodynamic disc simulation is required to verify this expectation.

Similarly, cosmic rays provide an additional source of pressure support against gravitational collapse, the magnitude of which may vary substantially across the interstellar medium~\citep[e.g.][]{2021ApJ...910..126S}. In particular, the measurement of an average effective pressure due to cosmic rays, via gamma-ray measurements of the cosmic ray energy density, depends on the detailed gas distribution and the gas densities at which cosmic rays accumulate, which remain uncertain. By comparison to isolated galaxy simulations that include cosmic ray pressure~\citep[e.g.][]{2012MNRAS.423.2374U,2016ApJ...824L..30P,2021ApJ...910..126S}, we do not expect that the introduction of cosmic rays would overwhelm the contribution of the internal molecular cloud velocity dispersion to the kpc-scale molecular gas pressure, and so we do not expect that it would substantially alter the results presented here.

\section{Conclusions} \label{Sec::conclusions}
In this work, we have studied the relationship between the scale-height of the molecular gas disc and the scale-dependent velocity dispersion of the molecular interstellar medium, using an isolated disc simulation of a Milky Way-like galaxy in the moving-mesh code {\sc Arepo}. We have found that:
\begin{enumerate}
\item The azimuthally-averaged vertical distribution of molecular gas volume densities is consistent with a Gaussian profile, as expected for a thin gas disc in a state of hydrostatic equilibrium within a vertically extended matter distribution.
\item The azimuthally-averaged scale-height of the molecular gas disc varies from $\sim 35$~pc to $\sim 50$~pc, and is constant with galactocentric radius $R$ for $R>5$~kpc. It does not display flaring behaviour.
\item If we assume that the molecular gas disc is in a state of hydrostatic equilibrium supported by the total kpc-scale molecular gas velocity dispersion, the resulting prediction for the molecular disc scale-height is an over-estimate of the true scale-height at all galactocentric radii $R$. The discrepancy increases from $\sim 15$~pc at $R \sim 3$~kpc up to $\sim 60$~pc at the edge of the galactic disc.
\item If the velocity dispersion between molecular cloud centroids is used instead of the kpc-scale velocity dispersion, the hydrostatic prediction is close to the true scale-height, with a maximum discrepancy of $\sim 15$~pc for clouds identified as surface density isocontours. This approach yields stable results within the range of map resolutions from $18$~pc to $150$~pc studied here.
This result is independent of the map resolution used for cloud identification.
\item When performing a parallel, pixel-by-pixel analysis, we find that result (iv) holds when the map resolution is comparable to or larger than the separations of discrete molecular clouds ($\sim 150$~pc). At higher resolution, the velocity dispersion between molecular-dominated sight-lines contains an increasing contribution from self-gravity, and so provides an increasingly-worse prediction for the scale-height.
\end{enumerate}

We conclude that the assumption of hydrostatic equilibrium can be applied to the molecular gas disc on galactic scales, to infer its vertical scale-height. However, due to the clumpiness of the molecular interstellar medium, this is not a hydrostatic equilibrium balancing the total gravitational force acting on the molecular gas and the total effective pressure within it. It is instead a hydrostatic equilibrium balancing the gravitational forces acting on giant molecular cloud centroids and the effective pressure resulting from the relative motions of these cloud centroids. The velocity dispersion inside giant molecular clouds does not contribute to the support of a Gaussian vertical distribution of gas volume densities.

Given the above results, observations of CO emission at high spatial resolution in external galaxies, made possible by instruments such as the Atacama Large Millimetre/Submillimetre Array (ALMA), will provide an opportunity to measure the molecular disc scale-height more accurately, using the velocity dispersion of the centroids of identified molecular clouds. In a companion paper, we will apply this technique to compute the molecular disc scale-height across a sub-set of the galaxies in the PHANGS sample.

\section*{Acknowledgements}
We thank an anonymous referee for a constructive report, which improved the clarity of Section~\ref{Sec::theory}. We thank Volker Springel for providing us with access to Arepo. SMRJ is supported by Harvard University through the ITC. The work of JS is partially supported by the Natural Sciences and Engineering Research Council of Canada (NSERC) through the Canadian Institute for Theoretical Astrophysics (CITA) National Fellowship. The work of JS is partially supported by the National Science Foundation (NSF) under Grants No.~1615105, 1615109, and 1653300. CDW acknowledges support from the Natural Sciences and Engineering Research Council of Canada and the Canada Research Chairs program. The work was undertaken with the assistance of resources and services from the National Computational Infrastructure (NCI; award jh2), which is supported by the Australian Government. We are grateful to Adam Leroy and Alyssa Goodman for helpful discussions.

\section*{Data Availability Statement}
The data underlying this article are available in the article and in its online supplementary material.



\bibliographystyle{mnras}
\bibliography{bibliography} 



\appendix
\section{Calculation of the molecular gas column density $\Sigma_{\rm H_2}$} \label{App::Sigma-H2-maps}
As noted in Section~\ref{Sec::sh-sims}, we identify molecular clouds in two-dimensional maps of the molecular gas column density $\Sigma_{\rm H_2}$. To calculate the molecular gas column density, we post-process the simulation output using the {\sc Despotic} model for astrochemistry and radiative transfer~\citep{Krumholz13a}. At the mass resolution of our simulation, the self-shielding of molecular hydrogen from the ambient UV radiation field cannot be accurately computed during run-time, so that the molecular hydrogen abundance is under-estimated by a factor $\sim 2$, requiring this value to be re-calculated in post-processing. Within {\sc Despotic}, the escape probability formalism is applied to compute the CO line emission from each gas cell according to its hydrogen atom number density $n_{\rm H}$, column density $N_{\rm H}$ and virial parameter $\alpha_{\rm vir}$, assuming that the cells are approximately spherical. In practice, the line luminosity varies smoothly with the variables $n_{\rm H}$, $N_{\rm H}$, and $\alpha_{\rm vir}$. We therefore interpolate over a grid of pre-calculated models at regularly-spaced logarithmic intervals in these variables to reduce computational cost. The hydrogen column density is estimated via the local approximation of~\cite{Safranek-Shrader+17} as $N_{\rm H}=\lambda_{\rm J} n_{\rm H}$, where $\lambda_{\rm J}=(\pi c_s^2/G\rho)^{1/2}$ is the Jeans length, with an upper limit of $T=40~{\rm K}$ on the gas cell temperature. The virial parameter is calculated from the turbulent velocity dispersion of each gas cell according to~\cite{MacLaren1988,BertoldiMcKee1992}. The line emission is self-consistently coupled to the chemical and thermal evolution of the gas, including carbon and oxygen chemistry~\citep{Gong17}, gas heating by cosmic rays and the grain photo-electric effect, line cooling due to ${\rm C}^+$, ${\rm C}$, ${\rm O}$ and ${\rm CO}$ and thermal exchange between dust and gas. We match the ISRF strength and cosmic ionisation rate to the values used in our live chemistry.

Having calculated values of the CO line luminosity for each simulated gas cell, we compute the CO-bright molecular hydrogen surface density as
\begin{equation}
\begin{split}
\Sigma_{\rm H_2}[{\rm M}_\odot{\rm pc}^{-2}] = &\frac{2.3 \times 10^{-29}{\rm M}_\odot({\rm erg~s}^{-1})^{-1}}{m_{\rm H}[{\rm M}_\odot]} \\
&\times \int^{\infty}_{-\infty}{\dd z^\prime \rho_{\rm g}(z^\prime) L_{\rm CO}[{\rm erg~s}^{-1}~{\rm H~atom}^{-1}]},
\end{split}
\end{equation}
where $\rho_{\rm g}(z)$ is the total gas volume density in ${\rm M}_\odot~{\rm pc}^{-3}$ at a distance $z$ (in ${\rm pc}$) from the galactic mid-plane. The factor of $2.3 \times 10^{-29}~{\rm M}_\odot~({\rm erg~s}^{-1})^{-1}$ combines the mass-to-luminosity conversion factor $\alpha_{\rm CO}=4.3~{\rm M}_\odot{\rm pc}^{-2}({\rm K}~{\rm kms}^{-1})^{-1}$ of~\cite{Bolatto13} with the line-luminosity conversion factor $5.31 \times 10^{-30}({\rm K~kms}^{-1}{\rm pc}^2)/({\rm erg~s}^{-1})$ for the CO $J=1\rightarrow 0$ transition at redshift $z=0$~\citep{SolomonVandenBout05}.


\bsp	
\label{lastpage}
\end{document}